\documentclass[
11pt,
a4paper,
oneside, 
headinclude,footinclude, 
BCOR5mm,
]{article}
\usepackage[utf8]{inputenc}
\usepackage{amsmath}
\usepackage{subcaption}
\usepackage{caption}
\usepackage{longtable}
\usepackage{amsthm}
\usepackage{color}
\usepackage{sectsty}
\usepackage[dvipsnames]{xcolor}
\usepackage{diagbox}
\usepackage{hyperref}
\usepackage{authblk}
\hypersetup{
    colorlinks=true, 
    linkcolor=blue, 
    urlcolor=red, 
    linktoc=all 
}

\usepackage{amssymb}
\usepackage{pagecolor,lipsum}
\usepackage{enumerate}
\usepackage{pdfpages}
\usepackage{xcolor}
\usepackage{tikzsymbols}
\usepackage{blkarray}
\usepackage{enumitem}
\usepackage{graphicx}
\usepackage{setspace}
\usepackage{soul}
\usepackage{enumitem}
\usepackage{soul}
\usepackage{MnSymbol}
\usepackage{mathtools}
\usepackage{bbm}
\usepackage{braket}
\usepackage[framemethod=tikz]{mdframed}
\usepackage{amsmath}
\usepackage[thinc]{esdiff}
\usepackage{array}
\usepackage{booktabs}
\DeclarePairedDelimiter\ceil{\lceil}{\rceil}

\DeclareMathOperator*{\argmax}{arg\,max}
\DeclareMathOperator*{\argmin}{arg\,min}
\usepackage[linesnumbered,ruled,vlined,english]{algorithm2e}
\usepackage{geometry}
 \usepackage{tgtermes}

 \let\oldpara\paragraph
\renewcommand{\paragraph}[1]{\vspace{-0.5cm}\oldpara{#1}}

\definecolor{mycolor}{rgb}{0.122, 0.435, 0.698}
\newmdenv[topline=false, bottomline=false, rightline=false, innerlinewidth=0.4pt, roundcorner=4pt,linecolor=mycolor,innerleftmargin=6pt,
innerrightmargin=6pt,innertopmargin=1pt,innerbottommargin=6pt]{mybox}

\newmdenv[backgroundcolor=gray!10, topline=false, bottomline=false, rightline=false, innerlinewidth=0.4pt, roundcorner=4pt,linecolor=black,innerleftmargin=6pt,
innerrightmargin=6pt,innertopmargin=3pt,innerbottommargin=6pt]{mybox2}

\newmdenv[backgroundcolor=blue!5, topline=false, bottomline=false, rightline=false, leftline=false, innerlinewidth=0.4pt, roundcorner=4pt,innerleftmargin=10pt,
innerrightmargin=10pt,innertopmargin=10pt,innerbottommargin=10pt]{mybox3}

\newcommand{\mc}[1]{\mathcal{#1}}
\newcommand{\mb}[1]{\mathbb{#1}}

\newcommand{\ts}[1]{\textsc{#1}}
\newcommand{\eqe}[1]{\textsc{EQE}}
\newcommand{\ep}[1]{\mathbb{E}}
\newcommand{\done}[1]{\textcolor{blue}{(\textbf{Done})}}

\DeclareMathAlphabet{\mymathbb}{U}{BOONDOX-ds}{m}{n}
\newcommand{\szero}[1]{\mymathbb{0}}
\newcommand{\sone}[1]{\mymathbb{1}}
\newcommand{\z}[1]{\mathbb{Z}}

\newcommand{\zt}[1]{\mathbb{Z}_T}

\newcommand{\np}{\textbf{NP}}

\newcommand{\G}{\mathcal{G}}
\newcommand{\Gv}{\mathcal{V}}
\newcommand{\Ge}{\mathcal{E}}
\newcommand{\A}{\mathcal{A}}
\newcommand{\W}{\mathcal{W}}

\newcommand{\assignment}{assignment}
\newcommand{\assignments}{assignments}

\newtheorem{theorem}{\textbf{Theorem}}[section]
\newtheorem{corollary}{\textbf{Corollary}}[theorem]
\newtheorem{claim}{\textbf{Claim}}[theorem]
\newtheorem{lemma}[theorem]{\textbf{Lemma}}
\newtheorem{definition}[theorem]{\textbf{Definition}}

\newtheorem{proposition}[theorem]{Proposition}
\newtheorem{observation}[theorem]{Observation}

\newcommand{\invSyDSc}[1]{(\ts{Inv-Thresh}, \ts{Closed-Nei})-SyDS}
\newcommand{\invSyDSo}[1]{(\ts{Inv-Thresh}, \ts{Open-Nei})-SyDS}

\newcommand{\invSDSc}[1]{(\ts{Inv-Thresh}, \ts{Closed-Nei})-SDS}
\newcommand{\invSDSo}[1]{(\ts{Inv-Thresh}, \ts{Open-Nei})-SDS}

\newcommand{\maxinte}[1]{\textsc{IM}}
\newcommand{\IoAgent}[1]{\texttt{IoA}}
\newcommand{\IoEdge}[1]{\texttt{IoE}}
\newcommand{\IoInter}[1]{\texttt{IoI}}
\newcommand{\IoDis}[1]{\texttt{IoD}}
\newcommand{\Io}[1]{\texttt{Io}}
\newcommand{\pla}[1]{\mathcal{P}}
\newcommand{\plas}[1]{\mathcal{P}^*}
\newcommand{\GvU}[1]{\mathcal{V}^{\texttt{U}}_{#1}}
\newcommand{\NU}[1]{\mathcal{N}^{\texttt{U}}_{#1}}
\newcommand{\Vto}[1]{\Tilde{\Gv{}}_{2-1}}
\newcommand{\Vot}[1]{\Tilde{\Gv{}}_{1-2}}

\definecolor{darkblue}{RGB}{0,0,76}

\title{\textbf{Assigning Agents to Increase Network-Based \\ Neighborhood Diversity}}

\author { \small
    Zirou Qiu,\textsuperscript{1,2}
    Andrew Yuan,\textsuperscript{2}
    Chen Chen,\textsuperscript{2}
    Madhav V. Marathe,\textsuperscript{1,2}
    S. S. Ravi,\textsuperscript{2,3} \\
    Daniel J. Rosenkrantz,\textsuperscript{2,3}
    Richard E. Stearns,\textsuperscript{2,3}
    Anil Vullikanti\textsuperscript{1,2}
}

{
\small
\affil[1]{\small Computer Science Dept., University of Virginia.}
\affil[2]{\small Biocomplexity Institute and Initiative, University of Virginia.}
\affil[3]{\small Computer Science Dept., University at Albany – SUNY.}
}

\date{}

\begin{document}
\maketitle

\begin{abstract}
\noindent
Motivated by real-world applications such as the allocation of public housing, we examine the problem of assigning a group of agents to vertices (e.g., spatial locations) of a network so that the {\em diversity level} is {\em maximized}. Specifically, agents are of two types (characterized by features), and we measure diversity by the number of agents who have at least one neighbor of a different type. This problem is known to be \np-hard, and we focus on developing approximation algorithms with provable performance guarantees. We first present a local-improvement algorithm for general graphs that provides an approximation factor of $1/2$. For the special case where the sizes of agent subgroups are similar, we present a randomized approach based on semidefinite programming that yields an approximation factor better than $1/2$. Further, we show that the problem can be solved efficiently when the underlying graph is treewidth-bounded and obtain a polynomial time approximation scheme (PTAS) for the problem on planar graphs. Lastly, we conduct experiments to evaluate the performance of the proposed algorithms on synthetic and real-world networks.

\smallskip
\noindent
\textbf{Conference version.} The conference version of the paper is accepted at \texttt{\textbf{AAMAS-2023}}: \href{https://dl.acm.org/doi/10.5555/3545946.3598690}{\textbf{Link}}. 
\end{abstract}

\section{Introduction}
Many countries have public housing initiatives that offer low-income individuals secure and affordable residences. Housing options are typically allocated by government agencies that involve a process of {\em assigning applicants to vacant apartments}~\cite{thakral2016public, glass2018remaking}. Given that the applicants often come from a variety of demographic groups, the spatial distribution of public housing partially shapes the demographic structure of local communities~\cite{skifter2016impact,gomez2019diversity}. The promotion and cultivation of integrated communities is an objective of contemporary societies. It has been shown that integration can improve a country's financial performance, reduce the disparity between demographic groups, and advance social prosperity in general~\cite{cheruvelil2014creating, mahadeo2012board,smith2000benefits}. 
Conversely, segregated neighborhoods widen the socioeconomic divide in the population. As noted by many social scientists, residential segregation remains a persistent problem that directly contributes to the uneven distribution of resources and limited life chances for some groups (e.g., \cite{rosen2021racial,tammaru2020relationship,van2020changing}).

\par In this work, we study the problem of promoting community integration (i.e., diversity) in the context of housing assignment. Indeed, public housing programs often take diversity into account. In Singapore, there are established policies to ensure that a certain ethnic quota must be satisfied for each project at the neighborhood level~\cite{debates52official}. In the U.S., cities like Chicago and New York also place emphasis on the value of having integrated communities~\cite{chicagodiversity,nycdiversity}. Nevertheless, {\em formal computational methods} for improving the level of integration in the housing assignment process have received limited attention. Motivated by the above considerations, we investigate the problem of public housing allocation from an algorithmic perspective and {\em provide systematic approaches to design assignment strategies that enhance community integration.}

Formally, we model a housing project as a graph $\G{} = (\Gv{}, \Ge{})$ where $\Gv{}$ is the set of vacant residences, and the edges in $\Ge{}$ represent proximity between residences. We are also given a set $\mc{A}$ of agents, representing the applicants to be assigned to the residences in $\Gv{}$. Agents are partitioned into {\em two} demographic subgroups: type-$\texttt{1}$ and type-\texttt{2}. 
Without loss of generality, we assume that the number of type-1 agents does not exceed the number of type-2 agents.
(We sometimes use the phrase ``minority agents'' for type-\texttt{1} agents.) We also assume that the number of vacant residences (i.e., $|\Gv|$) equals the number of agents. Our goal is to construct an assignment (bijective mapping) $\pla{}$ of residences to agents  that maximizes the {\em the integration level} of the layout of agents on $\G{}$.

\begin{figure}[h!]
  \centering
    \includegraphics[scale=0.4]{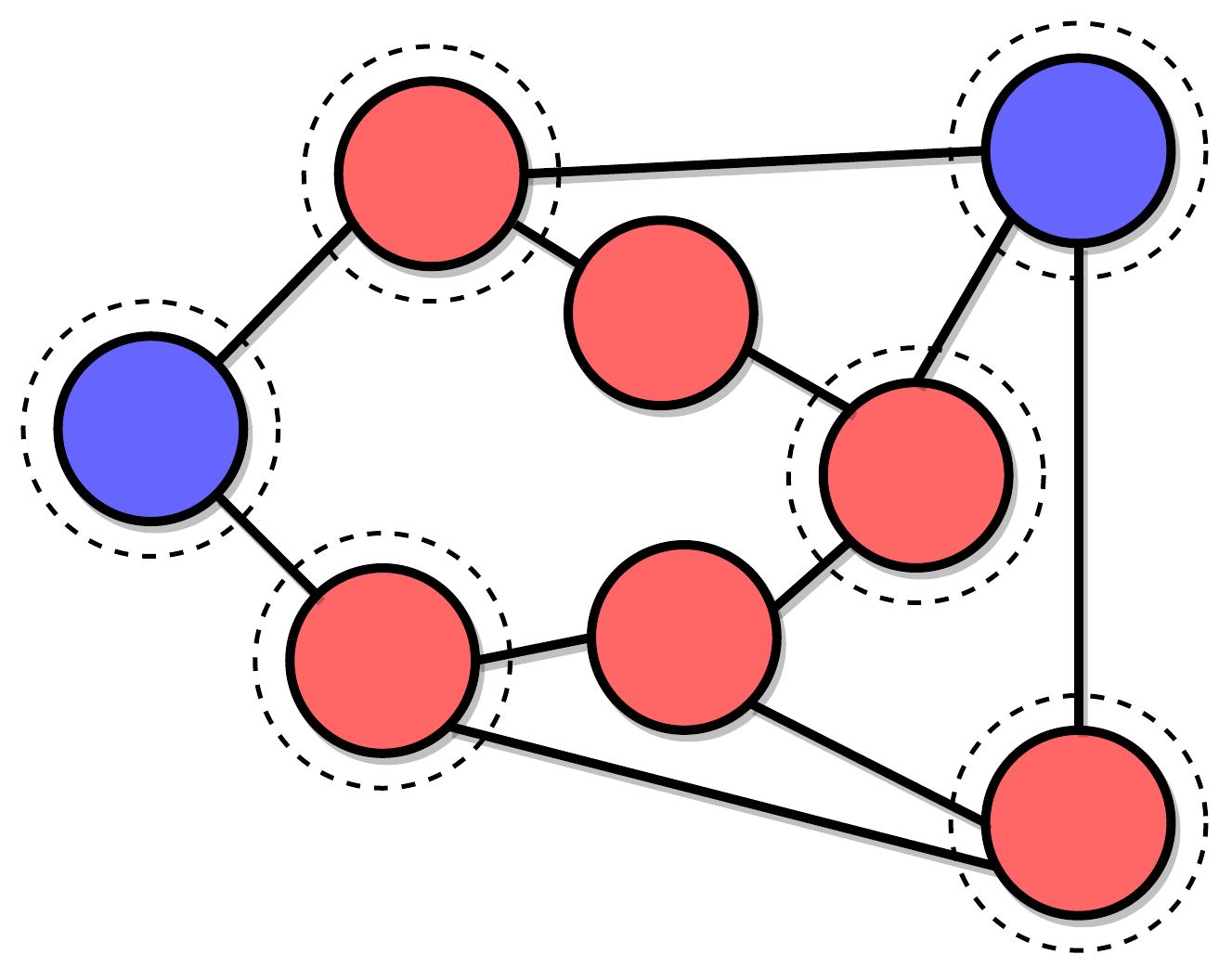}
    \caption{An example assignment of two type-\texttt{1} agents (blue) and six type-\texttt{2} agents (red) on a graph $\G{}$.
    Vertices with integrated agents are labeled by dashed circles. The index of integration for this assignment
    (ie., the number of integrated agents) is $6$.}
    \label{fig:example-index}
\end{figure}

To quantify the integration level of a given assignment $\mc{P}$, we use the {\em index of integration} (\IoAgent{}) metric proposed in~\cite{agarwal2020swap}.
This index is defined as the number of {\em integrated agents}, that is, agents with at least one neighbor of a different type in $\G{}$. 
An illustrative example is given in Fig.~\ref{fig:example-index}. 
We refer to the above assignment problem as \textsc{Integration Maximization - Index of Agent Integration} (\maxinte{}-\IoAgent{}). We note that this problem could also arise in other settings where integration is preferred, such as dormitory assignments for freshmen in universities~\cite{duncan2003empathy}. 

\par The problem of maximizing \IoAgent{} is known to be \np-hard~\cite{agarwal2020swap}. Nevertheless, the authors of~\cite{agarwal2020swap} did not  address approximation questions for the problem, as their focus is on game theoretic aspects of  \IoAgent{}. In this work, we focus on developing approximation algorithms with provable performance guarantees for \maxinte{}-\IoAgent{}. \textbf{Our main contributions are as follows.}
\smallskip
\noindent
\smallskip
\begin{itemize}[leftmargin=*,noitemsep,topsep=0pt]
    \item[-] \textbf{Approximation for general instances}. We present a {\em local-improvement algorithm} that guarantees a factor $1/2$ approximation. We further show that our analysis is tight by presenting an example that achieves this bound. While it is possible to derive an approximation for the problem using a general result in~\cite{buchbinder2014submodular}, the resulting performance guarantee is $0.356$, which is weaker than our factor of $1/2$.

    \item[-] \textbf{Improved approximation for special instances}. For the case when the number of type-1 agents is a constant fraction $\alpha$ of the total number of agents, $0 < \alpha \leq 1/2$, we present a semidefinite programming (SDP) based randomized algorithm that yields approximation ratios in the range $[0.516, 0.649]$ for $\alpha$ in the range $[0.403, 0.5]$. For example, when $\alpha = 0.45$, the ratio is $0.578$, and when $\alpha = 0.5$, the ratio is $0.649$.
    
    \item[-] \textbf{A polynomial time approximation scheme for planar graphs}. We present a {\em dynamic programming method} that solves \maxinte{}-\IoAgent{} efficiently on graphs with bounded treewidth. Using this result in conjunction with a technique due to Baker~\cite{baker1994approximation}, we obtain a {\em polynomial time approximation scheme} (PTAS) for the problem on planar graphs. For any fixed $\epsilon > 0$, the algorithm provides a performance guarantee of $1-\epsilon$. 
    
    \item[-] \textbf{Empirical analysis}. We study the empirical performance of the proposed local-improvement algorithm against baseline methods on both synthetic and real-world networks. Overall, we observe that the empirical approximation ratio of the proposed algorithm is much higher than $1/2$, which is our theoretical guarantee.
\end{itemize}

\section{Related Work}
\label{sec:related}

\paragraph{Integration in public housing.} Issues regarding segregation and the need for enhancing integration have been documented extensively in the social science literature (e.g., \cite{firebaugh2016blacks,Massey-Denton-1988,Lieberson-1980,Morgan-1983}). Many such works on segregation in social networks (e.g., \cite{Gretha-etal-2018,henry2011emergence}) stem from the pioneering models proposed by Schelling \cite{schelling1971dynamic} where agents move between vertices to improve their utility values. 
While Schelling's framework allows the study of
agent dynamics,  Benabbou et al.~\cite{benabbou2018diversity} study integration in public housing allocation from a {\em planning perspective}.
In particular, they formulate the setting as a weighted matching problem where the set of available houses is partitioned into blocks, and agents are assigned (by some central agency) to blocks. The goal of their problem is find an assignment that maximizes a utility measure while satisfying some {\em diversity} constraints. They establish the \np-hardness of the problem and present an approximation algorithm based on a result of Stamoulis~\cite{stamoulis2014approximation}. A number of other studies have also addressed integration in the context of public housing from a social science perspective (e.g., \cite{Pollack-etal-2014,kleit2001neighborhood,ludwig2012long,hananel2017central}).

\par The problem formulations and the algorithmic techniques used in Benabbou et al.~\cite{benabbou2018diversity} and in our work are significantly different. First, Benabbou et al.~\cite{benabbou2018diversity} examine a weighted matching problem. Their model does not use any network structure for the residences, whereas our work approaches the problem from a graph theoretic standpoint, with the underlying network playing an important role in the formulation. Further, the integration index studied in our work is defined w.r.t graph structures, whereas the measure used in~\cite{benabbou2018diversity} is based on constraints on the ethnicity quotas for blocks. More importantly, the goal of our work is to find an assignment that maximizes the integration level, whereas the goal in~\cite{benabbou2018diversity} is to maximize the overall utility of agents under a diversity constraint.

\paragraph{Integration indices.} Various indices to measure the level of integration in a population are surveyed in~\cite{Massey-Denton-1988}. However, most of those indices 
cannot be naturally extended to a network setting. The integration index \IoAgent{} considered in our work was proposed by Agarwal et al.~\cite{agarwal2020swap}\footnote{In Agarwal et al.~\cite{agarwal2020swap}, the index is called ``\textit{degree} of integration''. In our work, the term ``degree'' is used to denote the \textit{degree} of a vertex. We use the term ``\textit{index} of integration'' to denote the index proposed in~\cite{agarwal2020swap}.} in the context of the Schelling Game on networks, where agents can change locations to increase their utilities. 
Agarwal et al. explore several properties (e.g., the integration price of anarchy/stability) of the index from a game theoretic perspective. Further, they show that finding an assignment for which all agents are integrated (i.e., each agent has at least one neighbor of a different type) is NP-hard~\cite{agarwal2020swap}.

\paragraph{Approximation algorithms.} 
Our algorithm for general IM-\IoAgent{} is based on a local-improvement scheme.  A well-known problem for which a local-improvement algorithm provides an approximation guarantee of $1/2$ is the unweighted MaxCut problem~\cite{MU-2017}. We note that the analysis used to establish the performance guarantees of the local-improvement methods for MaxCut and IM-\IoAgent{} are substantially different. In particular, MaxCut has no cardinality constraints, and the objective is defined w.r.t edges. In contrast, IM-\IoAgent{} requires that a specified number of vertices be assigned to type-\texttt{1} agents, and the objective is defined w.r.t vertices. One can also formulate \maxinte{}-\IoAgent{} as a {\em non-monotone} submodular function maximization problem.
Since such a formulation requires a strict equality constraint (involving type-\texttt{1} agents), the best known
performance guarantee under the general {\em non-monotone} submodular maximization framework with such
a constraint is $0.356$~\cite{buchbinder2014submodular}.

\section{Problem Definition}
We study the problem of assigning {\em vertices} in a {\em graph} to a group of {\em agents}, such that the {\em integration level} of the resulting layout of agents in the graph is maximized. 
We begin with key notations and then define the integration maximization problem formally.

\paragraph{Graphs and agents.} Let $\G{} = (\Gv{}, \Ge{})$ be an undirected graph, where $\Gv{}$ is a set of vertices representing vacant residences, and $\Ge{}$ is a set of edges representing the proximity relationship between residences. Let $\A{}$ be the set of agents to be assigned to $\Gv{}$. The set of agents is divided into {\em two} demographic subgroups. Formally, $\mc{A}$ is partitioned into two subsets $\A{}_{1}$ and $\A{}_{2}$; we refer to agents in $\A{}_i$ as type $i$ agents, $i = \texttt{1}, \texttt{2}$. Let $k = |\A{}_{\texttt{1}}|$ denote the number of type-\texttt{1} agents, so $n-k$ is the number of type-\texttt{2} agents. Without loss of generality, let $k \leq n/2$, and we refer to $\A{}_{\texttt{1}}$ as the {\em minority subgroup}. Lastly, we assume that $|\Gv{}| = |\A|$; that is, the number of vertices is the same as the number of agents. 

\paragraph{Assignment.} An assignment is a mapping from vertices to agents. To simplify the proofs, we use an {\em equivalent definition} where an assignment is a mapping from vertices to agent types. In particular, an \assignment{} $\pla{} : \Gv{} \rightarrow \{\texttt{1}, \texttt{2}\}$ is a function that assigns an {\em agent type} to each vertex in $\Gv$, such that $k$ vertices are assigned type-\texttt{1} and $n - k$ vertices are assigned type-\texttt{2}. In such an assignment, a type-$i$ vertex is occupied by a type-$i$ agent, $i = \texttt{1}, \texttt{2}$. We remark that the above definition of an assignment is mathematically equivalent to defining an assignment to be a mapping from $\Gv{}$ to $\A{}$.

\paragraph{The index of integration.} We consider the integration index proposed in~\cite{agarwal2020swap} and apply it to our context. 

\smallskip

\begin{mybox2}
   \begin{definition}[\textbf{Index of agent-integration} (\IoAgent{})~\cite{agarwal2020swap}] Given an assignment $\pla{}$, an agent $x \in \mc{A}$ is \textbf{integrated} if $x$ has at least one neighbor in $\G{}$ whose type is different from that of $x$. Let $\mc{A}'$ be the set of integrated agents under $\mc{P}$. The index of agent-integration of $\mc{P}$ is then defined as the number of integrated agents in $\mc{A}$: 
\begin{equation}
    \IoAgent{}(\mathcal{P}) = |\mc{A}'|
\end{equation}
\end{definition} 
\end{mybox2}

\noindent
Equivalently, a vertex $u \in \Gv{}$ is \textit{integrated} under $\pla{}$ if the agent assigned to $u$ is integrated. Thus, we may also view the index as $\IoAgent{}(\mathcal{P}) = |\mc{V}'|$ where $\Gv{}'$ is the set of integrated vertices under $\pla{}$. These two definitions of $\IoAgent{}$ are mathematically equivalent. 

\paragraph{The optimization problem.} We now define the problem \maxinte{}-\IoAgent{}. 

\smallskip
\begin{mybox2}
    \begin{definition}[\textbf{\maxinte{}-\IoAgent{}}]
    Given a graph $\G{} = (\Gv{}, \Ge{})$, a set $\mc{A}$ of agents with $k$ type-\texttt{1} and $n-k$ type-\texttt{2} agents, find an assignment $\mathcal{P}$ such that $\IoAgent{}(\mathcal{P})$ is maximized.
    \end{definition}
\end{mybox2}

\noindent
We note that \maxinte{}-\IoAgent{} can be viewed as an optimization version of 2-\textit{weak} coloring~\cite{naor1993can}, where the number of vertices with each color is specified, and the number of properly colored vertices is maximized.

\section{Approximation for General Graphs}
\label{sec:general_graphs}
\maxinte{}-\IoAgent{} is \textbf{NP}-hard, as established in~\cite{agarwal2020swap}. In this section, we present a {\em local-improvement algorithm} for \maxinte{}-\IoAgent{} and show that the algorithm achieves a factor $1/2$ approximation for general graphs. For convenience in presenting the proofs, we consider an {\em \assignment{}} from the perspective of vertices rather than that of the agents. As stated earlier, these two definitions are equivalent.

\paragraph{The algorithm.} We start from a random \assignment{} $\mc{P}$. In each iteration of the algorithm, we find (if possible) a pair of type-\texttt{1} and type-\texttt{2} vertices such that swapping their types strictly increases the objective. In particular, let $u$ be a type-\texttt{1} vertex,  and $v$ be a type-\texttt{2} vertex. We swap the types of $u$ and $v$ (i.e., $u$ becomes type-\texttt{2} and $v$ becomes type-\texttt{1}) if and only if the resulting new \assignment{} $\mc{P}'$ has a strictly higher $\IoAgent{}$; that is, $\IoAgent{}(\mc{P}) < \IoAgent{}(\mc{P}')$. The algorithm terminates when no such swap can be made. The pseudocode is given 
in Algorithm~(\ref{alg:algo1-IoA}).

\begin{algorithm}
\caption{\texttt{Local-Improvement-\IoAgent{}}}
\label{alg:algo1-IoA}
\SetKwInOut{Input}{Input}
\SetKwInOut{Output}{Output}

\Input{A graph $\G{} = (\Gv{}, \Ge{})$, $k$, where $k \leq |\Gv{}|/2$}
\Output{An \assignment{} $\mc{P}$}

$\mc{P} \gets $ a random \assignment{} \& $\texttt{Updated} \gets \text{True}$

\While{$\texttt{Updated}$}
{
    $\texttt{Updated} \gets \text{False}$
    
    \For{$x \in \Gv{}_1(\pla{})$}
    {
        \For{$y \in \Gv{}_2(\pla{})$}
        {
            $\mc{P}' \gets$ the assignment where $\mc{P}'(x) = \mc{P}(y)$ and $\mc{P}'(y) = \mc{P}(x)$
            
            \If{$\IoAgent{}(\mc{P}') > \IoAgent{}(\mc{P})$}
            {
                $\mc{P} = \mc{P}'$, $\texttt{Updated} \gets \text{True}$ \& \textbf{break}
            }
        }
    }
}
\Return{$\mc{P}$}
\end{algorithm}

\subsection{Analysis of the algorithm}
\par Given a problem instance of \maxinte{}-\IoAgent{}, let $\mc{P}$ be a \underline{saturated} \assignment{}\footnote{An \assignment{} is {\em saturated} if no pairwise swap of types between a type-\texttt{1} and a type-\texttt{2} vertices can increase the objective.} returned by Algorithm~(\ref{alg:algo1-IoA}). Let $\plas{}$ be an optimal assignment that achieves the maximum objective, denoted by OPT. We assume that $\pla{} \neq \plas{}$. In this section, we show that $\IoAgent{}(\pla{}) \geq 1/2 \cdot \IoAgent{}(\plas{}) = 1/2 \cdot \textsc{OPT}$, thereby establishing a $1/2$ approximation. Due to the page limit, we sketch the proof here;  the full proof appears {\em in the appendix}.

Given an \assignment{} $\mathcal{P}$, which is a mapping from vertices to agent types, we call a vertex $v$ a {\em type-\texttt{1}} (or {\em type-\texttt{2}}) vertex if $\mathcal{P}(v) = 1$ (or $\mathcal{P}(v) = 2$). Let $\Gv{}_1(\mathcal{P})$ and $\Gv{}_2(\mathcal{P})$ denote the set of type-\texttt{1} and type-\texttt{2} vertices under $\mc{P}$. 
Let $\GvU{1}(\mathcal{P})$ and $\GvU{2}(\mathcal{P})$ denote the set of uncovered\footnote{Under an assignment, a vertex is ``covered'' if it is integrated and ``uncovered" otherwise.} type-\texttt{1} and type-\texttt{2} vertices under $\mc{P}$. 
For each vertex $u$, let $\NU{u}(\mathcal{P})$ denote the set of neighbors of $u$ that are uncovered under $\mc{P}$,
and let $\Gamma_u(\mathcal{P})$ denote the set of different-type neighbors of $u$ that are \textbf{uniquely} covered by $u$, i.e., $\Gamma_u(\mathcal{P})$ is the set of vertices $v$ such that $(i)$ $v$ is a neighbor of $u$, $(ii)$ the type of $v$ is different from the type of $u$, and $(iii)$ $v$ has no other neighbor whose type is the same as $u$'s type. 

\begin{mybox2}
\begin{observation}
The index $\IoAgent{}(\mc{P}) = n - |\GvU{1}(\mc{P})| - |\GvU{2}(\mc{P})|$.
\end{observation}
\end{mybox2}
\vspace{-0.5cm}

\noindent
We now consider the following mutually exclusive and collectively exhaustive cases of $\GvU{1}(\pla{})$ and $\GvU{2}(\pla{})$ under the saturated \assignment{} $\mc{P}$. We start with a simple case where all the type-\texttt{2} vertices under $\pla{}$ are integrated. 

\noindent
\underline{\textbf{Case 1}}: $\GvU{2}(\pla{}) = \emptyset$. 

\noindent
Under this case, all vertices in $\Gv{}_2(\pla{})$ are integrated which gives 
\begin{equation}
    \IoAgent{}(\mc{P}) \geq |\GvU{2}(\pla{})|  = n - k \geq \frac{1}{2} \cdot n \geq \frac{1}{2} \cdot \textsc{OPT}
\end{equation}

\noindent
The above case trivially implies that the
algorithm provides a $1/2$ approximation. We now look at the remaining case where $\GvU{2}(\pla{}) \neq \emptyset$. 

\noindent
\underline{\textbf{Case 2}}: $\GvU{2}(\pla{}) \neq \emptyset$. 

\noindent
Under this case, there exists at least one vertex in $\Gv{}_2(\pla{})$ that is not integrated. We first show that $\GvU{1}(\pla{})$ and $\GvU{2}(\pla{})$ cannot both be non-empty.

\begin{mybox2}
\begin{lemma}\label{lemma:ioa-1-main}
For a saturated \assignment{} $\mc{P}$, 
if $\GvU{2}(\pla{}) \neq \emptyset$, then $\GvU{1}(\pla{}) = \emptyset$.
\end{lemma}
\end{mybox2}

\begin{proof}(Sketch)
Let $y \in \GvU{2}(\pla{})$ be a vertex of type-\texttt{2} that is not integrated (i.e., all neighbors of $y$ are of type-\texttt{2}). For contradiction, suppose $\GvU{1}(\pla{}) \neq \emptyset$. Now let $x \in \GvU{1}(\pla{})$ be an non-integrated vertex of type-\texttt{1} whose neighbors are all of type-\texttt{1}. Let $\mc{P}'$ denote the \assignment{} where we switch the types between $x$ and $y$, that is, $\mc{P}'(x) = \mc{P}(y)= \texttt{2}$, $\mc{P}'(y) = \mc{P}(x) = \texttt{1}$, while the types of all other vertices remain unchanged. One can verify that $\IoAgent{}(\mc{P}') \geq \IoAgent{}(\mc{P}) + 2$, that is, switching the types of $x$ and $y$ increases the index \IoAgent{} by at least $2$. This implies the existence of an improvement move from $\mc{P}$, which contradicts the fact that $\mc{P}$ is saturated.  It follows that 
$\GvU{1}(\pla{}) = \emptyset$.  
\end{proof}

\noindent
Lemma~(\ref{lemma:ioa-1-main}) implies that under case 2 (i.e., $\GvU{2}(\pla{}) \neq \emptyset$), we have $\GvU{1}(\pla{}) = \emptyset$. We now consider the following two mutually exclusive and collectively exhaustive subcases under Case $2$ and show that the approximation factor under each subcase is $1/2$. 

\smallskip
\noindent
\underline{\textbf{Subcase 2.1:}} $\GvU{2}(\pla{}) \neq \emptyset$, and $\Gamma_x(\pla{}) \neq \emptyset, \; \; \forall x \in \Gv{}_1(\pla{})$, that is, for each type-\texttt{1} vertex $x \in \Gv{}_1(\pla{})$, there is at least one type-\texttt{2} neighbor of $x$ that is {\em uniquely} covered by $x$. 

\begin{figure}[!h]
  \centering
    \includegraphics[width=0.65\textwidth]{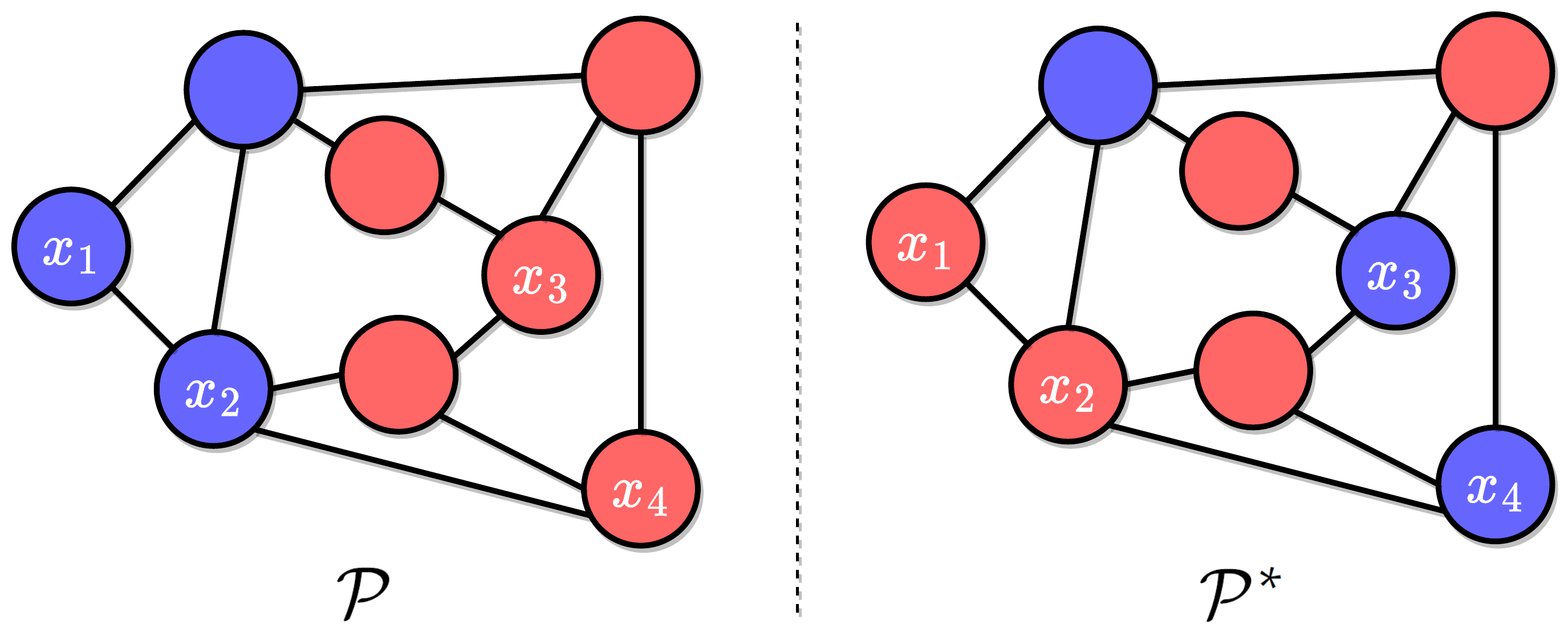}
    \caption{Two assignments $\pla{}$ and $\plas{}$ where type-\texttt{1} and type-\texttt{2} vertices are highlighted in blue and red, respectively. In this case, $\Vto{}= \{x_3, x_4\}$ and $\Vot{} = \{x_1, x_2\}$. We may then transform $\pla{}$ into $\plas{}$ by swapping types between the pair $(x_1, x_3)$ and between $(x_2, x_4)$. Note that this example is {\em only to demonstrate how $\Vto{}$ and $\Vot{}$ are defined}, as  $\pla{}$ cannot be a saturated \assignment{} returned by the algorithm.}
    \label{fig:p-ps-main}
\end{figure}

Suppose $\pla{} \neq \plas{}$, namely, for some $x \in \Gv{}$, $\pla{}(x) \neq \plas{}(x)$. Let $\Vto{} = \{v \in \Gv{} \; : \; \mc{P}(v) = 2, \mc{P}^*(v)  = 1\}$ be the set of vertices that are type-\texttt{2} under $\pla{}$, but are type-\texttt{1} under $\plas{}$. 
Analogously, let $\Vot{} = \{v \in \Gv{} \; : \; \mc{P}(v) = 1, \mc{P}^*(v) = 2\}$ be the set of vertices of type-\texttt{1} under $\pla{}$, but are of type-\texttt{2} under $\plas{}$. Observe that $|\Vto{}| = |\Vot{}|$. One may view $\mc{P}^*$ as the result of a transformation from $\pla{}$ under pairwise swaps of types between $\Vto{}$ and $\Vot{}$. An example is given in Figure~(\ref{fig:p-ps-main}).
We present a key lemma that bounds the difference between the objective values of $\pla{}$ and $\plas{}$.

\begin{mybox2}
\begin{lemma}[Subcase 2.1]\label{lem:subcase1-bound-main}
Let $\pla{}$ be a saturated \assignment{} under subcase 2.1, and let $\plas{}$ be an optimal \assignment{}. We have
\begin{align}
    \IoAgent{}(\plas{}) - \IoAgent{}(\pla{})
    &\leq \sum_{y \in \Vto{} \setminus \GvU{2}(\pla{})} |(\mc{N}(y) \cap \GvU{2}(\pla{})|\\  &+  \sum_{y \in \Vto{} \cap \GvU{2}(\pla{})} \left( |(\mc{N}(y) \cap \GvU{2}(\pla{})| + 1 \right).\nonumber
\end{align}
\end{lemma}
\end{mybox2}

\begin{proof}(Sketch)
Since $\pla{}$ is saturated, Lemma~(\ref{lemma:ioa-1-main}) implies that all type-\texttt{1} vertices under $\pla{}$ are integrated. Thus, the difference $\IoAgent{}(\plas{}) - \IoAgent{}(\pla{})$ is at most the number of type-\texttt{2} vertices that are integrated under $\plas{}$ but are {\em not} integrated under $\pla{}$.

\par Let $f: \Vot{} \rightarrow  \Vto{}$ be an arbitrary bijective mapping. We may regard $\plas{}$ as a result of the transformation from $\pla{}$ via pairwise swaps of types between vertices specified by $f$ (i.e., the type of $x \in \Vot{}$ is swapped with the type of $f(x) \in \Vto{}$). Observe that only vertices in $\GvU{2}(\pla{})$ that are adjacent to $\Vto{}$ (or within $\Vto{}$) under $\pla{}$ can be newly integrated under $\plas{}$ after swapping $\Vot{}$ with $\Vto{}$ (by the definition of $\GvU{2}(\pla{})$, vertices in $\Vot{}$ have no neighbors in $\GvU{2}(\pla{})$.). It follows that for each vertex $y \in \Vto{}$, at most $|(\mc{N}(y) \cap \GvU{2}(\pla{})|$ of its neighbors can become newly integrated after transforming from $\pla{}$ to $\plas{}$. Further, if also $y \in  \Vto{} \cap \GvU{2}(\pla{})$, $y$ itself could also be newly integrated after the swap. We then have
\begin{align}
    \IoAgent{}(\plas{}) - \IoAgent{}(\pla{}) &\leq |\bigcup_{y \in \Vto{}} \mc{N}(y) \cap \GvU{2}(\pla{})| + |\Vto{} \cap \GvU{2}(\pla{})| \nonumber\\
    &\leq \sum_{y \in \Vto{} \setminus \GvU{2}(\pla{})} |(\mc{N}(y) \cap \GvU{2}(\pla{})| \\ &+ \sum_{y \in \Vto{} \cap \GvU{2}(\pla{})} \left( |(\mc{N}(y) \cap \GvU{2}(\pla{})| + 1 \right)\nonumber
\end{align}
where the last inequality follows from the union bound.
\end{proof}

We now proceed to show that the difference between $\IoAgent{}(\plas)$  and $ \IoAgent{}(\pla{})$ established in Lemma~(\ref{lem:subcase1-bound-main}) is {\em at most} $\IoAgent{}(\pla{})$, thereby establishing $\IoAgent{}(\pla{}) \geq \frac{1}{2} \cdot \IoAgent{}(\plas{})$. Recall that for each vertex $x \in \Gv{}$, $\Gamma_x(\pla{})$ is the set of neighbors of $x$ whose types are different from $x$, and are uniquely covered by $x$ under $\pla{}$. By the definition of Subcase $2.1$, $\Gamma_x(\mathcal{P})$ is not empty for all $x \in \Gv{}_1(\pla{})$. We first argue that for any $y \in \GvU{2}(\pla{})$ and any $x \in \Gv{}_1(\pla{})$, we have $|\mc{N}(y) \cap \GvU{2}(\pla{})| \leq |\Gamma_x(\mathcal{P})|$.

\begin{mybox2}
\begin{lemma}[Subcase 2.1]\label{lemma:subcase2.1-1-main}
Given a saturated \assignment{} $\pla{}$, for any $y \in \GvU{2}(\pla{})$ and $x \in \Gv{}_1(\pla{})$, we have 
$$|\mc{N}(y) \cap \GvU{2}(\pla{})| \leq |\Gamma_x(\mathcal{P})|.$$
\end{lemma}
\end{mybox2}

\begin{proof}(Sketch)
Given that $y$ is not integrated under $\pla{}$, note that $x$ and $y$ {\em cannot} be adjacent. 
Since $\pla{}$ is a saturated \assignment{}, if the types of $x$ and $y$ are to be swapped,
the number of newly integrated vertices would be {\em at most} the number of newly non-integrated vertices. Further, one can verify that the number of vertices that are newly integrated is at least $|\mc{N}(y) \cap \GvU{2}(\pla{})| + 1$, and the number of vertices that are newly non-integrated is at most $|\Gamma_x(\mathcal{P})| + 1$. Since $\mc{P}$ is saturated, it follows that $|\mc{N}(y) \cap \GvU{2}(\pla{})| \leq |\Gamma_x(\mathcal{P})|$. This concludes the proof.
\end{proof}

\noindent
We now establish a bound on the size of $\mc{N}(y) \cap \GvU{2}(\pla{})$ for $y \in \Gv{}_2(\pla{}) \setminus \GvU{2}(\pla{})$ and $x \in \Gv{}_1(\pla{})$.

\begin{mybox2}
\begin{lemma}[Subcase 2.1]\label{lemma:subcase2.1-2-main}
Given a saturated \assignment{} $\pla{}$, for any $y \in \Gv{}_2(\pla{}) \setminus \GvU{2}(\pla{})$ and any $x \in \Gv{}_1(\pla{})$, we have 
$$|\mc{N}(y) \cap \GvU{2}(\pla{})| \leq |\Gamma_x(\mathcal{P})| + 1$$
\end{lemma}
\end{mybox2}

\begin{proof}(Sketch)
    We partition $\Gv{}_2(\pla{}) \setminus \GvU{2}(\pla{})$ into two subsets $\mc{B}$ and $\mc{C}$, as follows. Subset $\mc{B}$ is the set of integrated type-\texttt{2} vertices whose neighbors are all integrated under $\mc{P}$, i.e., $\mc{B} = \{y \in \Gv{}_2(\pla{}) \setminus \GvU{2}(\pla{}) \; : \; \mc{N}(y) \cap \GvU{2}(\pla{}) = \emptyset\}$. Subset $\mc{C}$, the complement of $\mc{B}$, is the set of integrated type-\texttt{2} vertices with at least one non-integrated neighbor under $\pla{}$, i.e., $\mc{C} = \{y \in \Gv{}_2(\pla{}) \setminus \GvU{2}(\pla{}) \; : \; \mc{N}(y) \cap \GvU{2}(\pla{}) \neq \emptyset\}$. The lemma clearly holds if $y \in \mc{B}$. Further, we show that for the case when $y \in \mc{C}$, no type-\texttt{1} neighbors of $y$ is uniquely covered by $y$ under $\pla{}$ (i.e., $\Gamma_y(\mathcal{P}) = \emptyset$). Further, suppose $y \in \mc{C}$, consider an objective non-increasing move from $\pla{}$ where we swap the types between $x$ and $y$. If $y$ is a neighbor of $x$ under $\mc{P}$, one can verify that the the maximum loss is $|\Gamma_x(\mathcal{P})|$ and the minimum gain is $|\mc{N}(y) \cap \GvU{2}(\pla{})|$. Thus

\begin{equation}
    |\mc{N}(y) \cap \GvU{2}(\pla{})| \leq |\Gamma_x(\mathcal{P})|
\end{equation}
On the other hand, if $y$ is \textbf{not} a neighbor of $x$ under $\mc{P}$, one can verify that the maximum loss is $|\Gamma_x(\mathcal{P})| + 1$ and the minimum gain is $|\mc{N}(y) \cap \GvU{2}(\pla{})|$. Thus
\begin{equation}
    |\mc{N}(y) \cap \GvU{2}(\pla{})| \leq |\Gamma_x(\mathcal{P})| + 1
\end{equation}
This concludes the proof.
\end{proof}

\noindent
We are now ready to establish $\IoAgent{}(\pla{}) \geq \frac{1}{2} \cdot \IoAgent{}(\plas{})$ under Subcase 2.1. 

\begin{mybox2}
\begin{lemma}[Subcase 2.1]\label{lem:subcase2.1-final-main}
Suppose $\GvU{2}(\pla{}) \neq \emptyset$ and $\Gamma_x(\pla{}) \neq \emptyset, \forall x \in \Gv{}_1(\pla{})$, we have 
$\IoAgent{}(\pla{}) \geq \frac{1}{2} \cdot \IoAgent{}(\plas{})$
where $\plas{}$ is an optimal assignment that gives the maximum objective.
\end{lemma}
\end{mybox2}

\begin{proof} (Sketch)
Note that $\Vto{}$ is a subset of $\Gv{}_2(\pla{})$. Further, 
Observe that $\Gamma_x(\mathcal{P})$ are disjoint for different vertices $x \in \Gv{}_1(\pla{})$. Now, by Lemma~(\ref{lem:subcase1-bound}) to and~(\ref{lemma:subcase2.1-2-main}), We have
\begin{align}
    &\IoAgent{}(\plas{}) - \IoAgent{}(\pla{}) \nonumber\\ 
    &\leq \left(\sum_{y \in \Vto{}} |\Gamma_{f^{-1}(y)}(\mathcal{P})|\right) +  |\Vto{}| \nonumber\\
    &\leq |\Gv{}_2(\pla{}) \setminus \GvU{2}(\pla{})| + |\Gv{}_1(\pla{})|\label{ineq:subcase2.1-main}\\
    &\leq \IoAgent{}(\pla{})\nonumber
\end{align}
\noindent
where Ineq~(\ref{ineq:subcase2.1-main}) follows from $|\Vto{}| = |\Vot{}| \leq |\Gv{}_1(\pla{})|$ and $\left(\sum_{y \in \Vto{}} |\Gamma_{f^{-1}(y)}(\mathcal{P})|\right) \leq |\Gv{}_2(\pla{}) \setminus \GvU{2}(\pla{})|$.
\end{proof}

We now have shown that, if $\GvU{2}(\pla{}) \neq \emptyset$ and $\Gamma_x(\pla{}) \neq \emptyset, \forall x \in \Gv{}_1(\pla{})$, the algorithm gives a $1/2$ approximation. We proceed to the final subcase.

\smallskip
\noindent
\underline{\textbf{Subcase 2.2:}} $\GvU{2}(\pla{}) \neq \emptyset$, and  $\Gamma_x(\pla{}) = \emptyset, \exists x \in \Gv{}_1(\pla{})$,
that is, there exists at least one type-\texttt{1} vertex $x \in \Gv{}_1(\pla{})$ such that for each type-\texttt{2} neighbor $y$ of $x$, $y$ is adjacent to at least one type-\texttt{1} vertex {\em other than $x$}.

\begin{mybox2}
\begin{lemma}[Subcase 2.2]\label{lemma:ioa-2-main}
Under subcase 2.2, for each non-integrated type-\texttt{2} vertex $y \in \GvU{2}(\pla{})$, all type-\texttt{2} neighbors of $y$ are integrated (i.e., $\mc{N}(y) \cap \GvU{2}(\pla{}) = \emptyset$) under $\pla{}$. That is, the vertices in $\GvU{2}(\pla{})$ form an \textbf{independent set} of $\mc{G}$.
\end{lemma}
\end{mybox2}

\begin{proof}(Sketch)
 Given such a $x \in \Gv{}_1(\pla{})$ defined in Subcase 2.2, for contradiction, suppose there exists a non-integrated type-\texttt{2} vertex $y \in \GvU{2}(\pla{})$ such that at least one type-\texttt{2} neighbor, denoted by $y' \in \mc{N}(y)$, of $y$ is not integrated under $\mc{P}$ (note that all neighbors of $y$ are of type-\texttt{2} since $y$ is not integrated). Now consider a new \assignment{} $\mc{P}'$ where we switch the types between $x$ and $y$. One can verify that $\IoAgent{}(\mc{P}') \geq \IoAgent{}(\mc{P}) + 1$, that is, after the switch, the index \IoAgent{} would increase by at least $1$. This implies the existence of an improvement move from $\mc{P}$, which contradicts $\mc{P}$ being a saturated \assignment{}. Thus, no such a non-integrated type-\texttt{2} vertex $y'$ of $y$ can exist.
\end{proof}

\noindent
Observe that $\IoAgent{}(\mc{P}) = (n - |\GvU{2}(\pla{})|)$. Using Lemma~(\ref{lemma:ioa-2}), we now argue that the size of $\GvU{2}(\pla{})$ {\em cannot be too large}. 

\begin{mybox2}
\begin{lemma}[Subcase 2.2]\label{lemma:ioa-3-main}
Under Subcase 2.2,
\begin{equation}
    |\GvU{2}(\pla{})| \leq \frac{n}{2}
\end{equation}
\end{lemma}
\end{mybox2}

\begin{proof}(Sketch)
Let $\mc{Y} \vcentcolon= \{y \in \Gv{}_2(\pla{}) \setminus \GvU{2}(\pla{}) \; : \; \mc{N}(y) \cap \GvU{2}(\pla{}) \neq \emptyset\}$ be the set of type-\texttt{2} {\em integrated} vertices whose has at least one non-integrated type-\texttt{2} neighbor. We first note that $\Gamma_y(\mathcal{P})$ (if not empty) are mutually disjoint for different $y \in \mc{Y}$. It follows that $\IoAgent{}(\mc{P}) \geq |\mc{Y}| + \sum_{y \in \mc{Y}} |\Gamma_y(\mathcal{P})|$. Suppose we switch the types between such a vertex $x$ and a vertex $y \in \mc{Y}$, and let $\mc{P}'$ denote the resulting new \assignment{}. One can verify that the maximum loss of objective after the swap is $|\Gamma_y(\mathcal{P})| + 1$, whereas the minimum gain is $|\mc{N}(y) \cap \GvU{2}(\pla{})|$. Since $\mc{P}$ is a saturated \assignment{} returned by the algorithm, we must have $\texttt{IoA}(\mc{P}) \geq \texttt{IoA}(\mc{P}')$. Therefore, $|\mc{N}(y) \cap \GvU{2}(\pla{})| \leq |\Gamma_y(\mathcal{P})| + 1, \; \forall y \in \mc{Y}$. Overall, we have that
\begin{align}
    |\GvU{2}(\pla{})| &= |\bigcup_{y \in \mc{Y}} \mc{N}(y) \cap \GvU{2}(\pla{})| \\ 
    &\leq |\Gv{}_1(\pla{})| + |\mc{Y}| \\
    &\leq |\Gv{}_1(\pla{})| + |\Gv{}_2(\pla{}) \setminus \GvU{2}(\pla{})|\\
    &= n - |\GvU{2}(\pla{})|
\end{align}
It immediately follows that $|\GvU{2}(\pla{})| \leq \frac{n}{2}$. 
\end{proof}

\noindent
Lastly, Since $\IoAgent{}(\mc{P}) = n - |\GvU{2}(\pla{})|$, by Lemma~(\ref{lemma:ioa-3-main}), we have 
$$
\IoAgent{}(\mc{P}) = n - |\GvU{2}(\pla{})| \geq \frac{1}{2} \cdot n \geq \frac{1}{2} \cdot \IoAgent{}(\plas{})
$$
thereby establishing a $1/2$ approximation for Subcase 2.2. Overall, we have shown that a saturated \assignment{} $\pla{}$ returned by Algorithm~(\ref{alg:algo1-IoA}) gives a $1/2$-approximation for \maxinte{}-\IoAgent{}. 
Thus:

\begin{mybox2}
\begin{theorem}\label{thm-1/2main-main}
Algorithm~(\ref{alg:algo1-IoA}) gives a $\frac{1}{2}$-approximation for \maxinte{}-\IoAgent{}.
\end{theorem}
\end{mybox2}

\paragraph{Analysis is tight.} We present a class of problem instances where the approximation ratio of the solution produced by Algorithm~(\ref{alg:algo1-IoA}) can be arbitrarily close to $1/2$. Therefore, the ratio $1/2$ in the statement of Theorem~(\ref{thm-1/2main}) cannot be improved, so {\em our analysis is tight.} The proof appears in the Appendix.

\begin{mybox2}
\begin{proposition}
For every $\epsilon > 0$, there exists a problem instance of \maxinte{}-\IoAgent{} for which there is a saturated \assignment{} $\pla{}$ such that $\IoAgent{}(\pla{}) \leq (\frac{1}{2} + \epsilon) \cdot \textsc{OPT}$.
\end{proposition}
\end{mybox2}

\section{Subgroups With Similar Sizes}
\label{sec:sdp_results}

In this section, we study the problem under the scenario where the number of type-\texttt{1} agents is a constant fraction of the total number of agents, that is, $k = \alpha \cdot n$ for some constant $ 0 \leq \alpha \leq 1/2$. We refer to this problem as $\alpha n$-\maxinte{}-\IoAgent{}. For example, $\alpha = 1/2$ represents the {\em bisection} constraint. We first show that $\alpha n$-\maxinte{}-\IoAgent{} remains computationally intractable. The proof appears in the Appendix.
\begin{mybox2}
\begin{theorem}
The problem $\alpha n$-\maxinte{}-\IoAgent{} is \textbf{NP}-hard.
\end{theorem}
\end{mybox2}

\subsection{A semidefinite programming approach}
We now present an approximation algorithm for $\alpha n$-\maxinte{}-\IoAgent{} based on semidefinite programming (SDP) relaxation~\cite{goemans1995improved}. The overall scheme is inspired by the work of Frieze and Jerrum~\cite{frieze1997improved} on the Max-Bisection problem. Given a graph $\G{} = (\Gv{}, \Ge{})$, each vertex $i \in \Gv{}$ has a binary variable $x_i \in \{-1, 1\}$ such that $x_i = -1$ if $i$ is of type-\texttt{1}, and $x_i = 1$ if $i$ is of type-\texttt{2}. 
First, we observe that a valid quadratic program (QP) is:
$\text{maximize}
\sum_{i \in \Gv{}} \; \max_{j \in \mc{N}(i)} \; \{ \frac{1 - x_i x_j}{2} \}$
\text{s.t.} 
$\sum_{i < j} x_i x_j = \frac{(1-2\alpha)^2 \cdot n^2 - n}{2}$. 
It can be verified that the following SDP is a relaxation of the QP:
\noindent
\begin{align*}
\text{SDP}: \;\;\;\;\; \text{maximize} \quad
&\sum_{i \in \Gv{}} \; \max_{j \in \mc{N}(i)} \; \{ \frac{1 - \vec{y}_i \cdot \vec{y}_j}{2} \} \\
\text{s.t.} \quad
& \sum_{i < j} \vec{y}_i \cdot \vec{y}_j \leq \frac{(1-2\alpha)^2 \cdot n^2 - n}{2}\\
& \vec{y}_i \cdot \vec{y}_i = 1, \;\;\;\;\; \forall i \in V
\end{align*}

\paragraph{Main idea of the algorithm and analysis.}
Our algorithm involves two steps. We elaborate on these steps and the analysis below. 
\begin{enumerate}
\item 
The SDP solution $\vec{y}_i, i=1,\ldots,n$ is not a feasible integral solution. So we round it to get a partition $(\Gv{}_1, \Gv{}_2)$ using the {\em hyperplane rounding method}~\cite{goemans1995improved} approach. We show that the expected number of integrated vertices is $\Omega(OPT_{SDP})$, where $OPT_{SDP}$ is the value of the SDP solution.
\item Note that $\{\Gv{}_1, \Gv{}_2\}$ need not be a valid $(\alpha n, (1-\alpha)n)$-partition, so we fix it by moving $|\Gv{}_1| - \alpha n$ nodes from $\Gv{}_1$ to the other side. We present a greedy strategy that picks a vertex to remove from $\Gv{}_1$ at each step, which does not decrease the overall $\IoAgent{}$ significantly. To achieve the overall guarantees, we run the rounding and size adjustment step multiple times and take the best solution. 
\end{enumerate}

\vspace{-0.1cm}
\paragraph{First step: Round the SDP.} Let $\{\vec{y}_1, ..., \vec{y}_n\}$ be an  optimal solution to the SDP; let $\textsc{OPT}_{SDP}$ be the objective value of the SDP. We  round the SDP solution to a partition $\{\Gv{}_1, \Gv{}_2\}$ of the vertex set such that vertices in $\Gv{}_i$ are of type-$i$, $i = \texttt{1}, \texttt{2}$ by applying Goemans and Williamson's {\em hyperplane rounding method}~\cite{goemans1995improved}. In particular, we draw a random hyperplane thought the origin with a normal vector $r$, and then $\Gv{}_1 = \{i : \vec{y}_i \cdot r \geq 0\}$ and $\Gv{}_2 = \{i : \vec{y}_i \cdot r < 0\}$.  

\par Consider an \assignment{} $\pla{}$ generated by the above rounding method (i.e., vertices in $\Gv{}_i$ are assigned to type-$i$). Let $f(\Gv{}_1) : 2^{\Gv{}} \rightarrow \mathbb{N}$ be the number of integrated vertices under $\pla{}$. We establish the following lemma. A detailed proof appears in the Appendix.

\begin{mybox2}
\begin{lemma}\label{lemma:sdp-quality-main}
$\mathbb{E}[f(\Gv{}_1)] \geq \alpha_{GW} \cdot \textsc{OPT}_{SDP}$, where $\alpha_{GW} \geq 0.878567$.
\end{lemma}
\end{mybox2}

\begin{proof}(Sketch)
We first establish that 
$$\Pr[i \text{ is integrated}] \geq \max_{j \in \mc{N}(i)} \{ \; \frac{\arccos{} (\vec{y}_i \cdot \vec{y}_j)}{\pi} \; \}$$
for any vertex $i$. Further, as shown in~\cite{goemans1995improved}, $\arccos{}(z) / \pi \geq \alpha_{GW} \cdot (1 - z) / 2$ for real $z \in [-1, 1]$. Thus,
\begin{align}
    \mb{E}[f(\Gv{}_1)] &\geq \sum_{i \in \Gv{}} \max_{j \in \mc{N}(i)} \{ \; \frac{\arccos{} (\vec{y}_i \cdot \vec{y}_j)}{\pi} \; \}\\
    &\geq \alpha_{GW} \cdot \sum_{i \in \Gv{}} \max_{j \in \mc{N}(i)} \{\frac{1 - \vec{y}_i \cdot \vec{y}_j}{2}\}\\
    &\geq \alpha_{GW} \cdot \textsc{OPT}_{SDP}
\end{align}
This concludes the proof.
\end{proof}

\paragraph{Second step: Fix the size.} In the previous step, we have shown that given a partition $\{\Gv{}_1, \Gv{}_2\}$ resulting from hyperplane rounding, if all vertices in $\Gv{}_1$ are of type-\texttt{1}, and all vertices in $\Gv{}_2$ are of type-\texttt{2}, then the expected number of integrated vertices is at least $\alpha_{GW}$ of the optimal. However, the partition is not necessarily an $(\alpha n, 1- \alpha) n$-partition. Thus, we present an algorithm to move vertices from one subset to another such that $(i)$ the resulting new partition is an $(\alpha n, (1- \alpha) n)$-partition, and $(ii)$ the objective does not decrease ``too much'' after the moving process.

\noindent
\textbf{Algorithm 2:} \texttt{Fix-the-Size}. Without losing generality, suppose $|\Gv{}_1| \geq \alpha n$. Overall, our algorithm consists of $T = |\Gv{}_1| - \alpha n$ iterations, and in each each iteration, we move a vertex $i \in \Gv{}_1$ to $\Gv{}_2$. Specifically, let $\Gv{}_1^{(t)}$ be the subset at the $t$th iteration, with $\Gv{}_1^{(0)} = \Gv{}_1$. To obtain $\Gv{}_1^{(t+1)}$, we choose $i \in \Gv{}_1^{(t)}$ to be a vertex that maximizes $f(\Gv{}_1^{(t)} \setminus \{i\}) - f(\Gv{}_1^{(t)})$, and the move $i$ to the other subset. Lemma~(\ref{lem:average-increase-main}) below establishes the performance of Algorithm (2); detailed proof appears in the Appendix. 

\begin{mybox2}
\begin{lemma}\label{lem:average-increase-main}
We have 
\begin{equation}
    \frac{f(\Gv{}_1^{(T)})}{|\Gv{}_1^{(T)}|} ~\geq~ \frac{f(\Gv{}_1)}{|\Gv{}_1|}
\end{equation}
where $\Gv{}_1^{(T)}$,  with $T = |\Gv{}_1| - \alpha n$, is returned by Algorithm (2).
\end{lemma}
\end{mybox2}

\paragraph{The final algorithm.} We have defined the two steps (i.e., $(i)$ {\em round the SDP} and $(ii)$ {\em fix the sizes of the two subsets}) needed to obtain a feasible solution for the problem. Let $\epsilon \geq 0$ be a small constant, and let $L = \ceil{\log_{a}(\frac{1}{\epsilon})}$ where $a = [(1 + \beta) - (1 - \epsilon) 2 \alpha_{GW}] / (1 + \beta - 2 \alpha_{GW})$, $\beta = 1/(4(\alpha - \alpha^2))$. Note that $L$ is a constant w.r.t. $n$. The final algorithm consists of $L$ iterations, where each iteration performs the two steps defined above. This gives us $L$ feasible solutions. The algorithm then outputs a solution with the highest objective among the $L$ feasible solutions. 

\begin{mybox2}
\begin{theorem}
    The final algorithm gives a factor 
    $$\frac{\alpha \left( (1 - \epsilon) \cdot 2\alpha_{GW} - \frac{\gamma-\gamma^2}{\alpha-\alpha^2} \right)}{\gamma} \cdot (1 - \epsilon)$$ 
    approximation w.h.p. where $\alpha_{GW} \geq 0.878567$, $\epsilon \geq 0$ is an arbitrarily small positive constant, $\alpha = k/n$ is the fraction of minority agents in the group, and $\gamma = \sqrt{\alpha (1 - \alpha) (1 - \epsilon) \cdot 2 \alpha_{GW}}$.
\end{theorem}
\end{mybox2}

For small enough $\epsilon$, say $\epsilon = 10^{-3}$, the approximation ratio is greater than $1/2$ for $\alpha$ in range $[0.403, 0.5]$. For example, $\alpha = 0.45$ gives a ratio of $0.5781$, and $\alpha = 0.5$ gives a ratio of $0.6492$.

\section{Tree-width Bounded Graphs and Planar Graphs}
In this section, we show that \maxinte{}-\IoAgent{} can be solved in polynomial time on treewidth bounded graphs. Using this result, we obtain a {\em polynomial time approximation scheme} (PTAS) for the problem on planar graphs.

\subsection{A dynamic programming algorithm for treewidth bounded graphs}
The concept {\em treewidth} of a graph was introduced in the work of Robertson and Seymour~\cite{robertson1986graph}. 
Many graph problems that are \np-hard are
known to be solvable in polynomial time when the
underlying graphs have bounded treewidth. In this section, we present a polynomial time dynamic programming algorithm for  \maxinte{}-\IoAgent{} for the class of treewidth bounded graphs. We refer readers to the Appendix for the definition of a tree decomposition and treewidth. 

\noindent
\textbf{Dynamic programming setup.} Given an instance of \maxinte{}-\IoAgent{} with graph $\G = (\Gv{}, \Ge{})$ and the number $k$ of minority agents, let $\mc{T} = (\mc{I}, \mc{F})$ be a tree decomposition of $\G{}$ with treewidth $\sigma$. For each $\mc{X}_i \in \mc{I}$, let $\mc{Y}_i$ be the set of vertices in the bags in the subtree rooted at $\mc{X}_i$. Let $\G{}[\mc{Y}_i]$ denote the subgraph of $\G{}$ induced on $\mc{Y}_i$. For each bag $\mc{X}_i$, we define an array $H_i$ to keep track of the optimal objectives in $\G{}[\mc{Y}_i]$. In particular, let $H_i(S, S', \gamma)$ be the optimal objective value for $\G{}[\mc{Y}_i]$ such that $(i)$ vertices in the subset $S \subseteq \mc{X}_i$ are of type-\texttt{1} and vertices in $\mc{X}_i \setminus S$ are of type-\texttt{2}; $(ii)$ vertices in $S' \subseteq \mc{X}_i$ are to be {\em treated as integrated}; 
$(iii)$ $\G{}[\mc{Y}_i]$ has  a total of $\gamma$ type-\texttt{1} vertices and $|\mc{Y}_i| - \gamma$ type-\texttt{2} vertices. For space reasons,  the update scheme for $H_i$ for each bag $\mc{X}_i$ and the proof of correctness appear in the appendix. 

\begin{mybox2}
\begin{theorem}\label{thm:twbounded}
\maxinte{}-\IoAgent{} can be solved in polynomial time on treewidth bounded graphs. 
\end{theorem}
\end{mybox2}

\subsection{PTAS for planar graphs}
Based on the result in \cite{agarwal2020swap}, it is easy to verify that \maxinte{}-\IoAgent{} remains hard on planar graphs. Given a planar graph $\mc{G}$ and for any fixed $\epsilon > 0$, based on the technique introduced in~\cite{baker1994approximation}, we present a {\em polynomial time approximation scheme} that achieves a $(1 - \epsilon)$ approximation for \maxinte{}-\IoAgent{}. 

\paragraph{PTAS Outline.} Let $q = 2 \cdot \ceil{1/\epsilon}$. We start with a plane embedding of $\G{}$, which partitions the set of vertices into $\ell$ layers for some integer $\ell \leq n$. Let $\Gv{}_i$ be the set of vertices in the $i$th layer, $i = 1, ..., \ell$.  For each $r = 1, ..., q$, observe that we may partition the vertex set into $t+1$ subsets, where $t = \ceil{(\ell - r) / q}$, such that the $(i)$ the first subset $\mc{W}_{(1, r)}$ consists of the first $r$ layers, $(ii)$ the last subset $\mc{W}_{(t+1, r)}$ consists of the last $\left((l - r) \mod q \right)$ layers, and $(iii)$ each $i$th subset $\mc{W}_{(i, r)}$ in the middle contains $q$ layers in sequential order. Let $\mc{W}_r = \{\mc{W}_{(1, r)}, ..., \mc{W}_{(t+1, r)}\}$ be such a partition. Let $\G{}_{(i,r)}$ be the subgraph induced on $\mc{W}_{(i, r)}$, $i = 1. ,,, t+1$. It is known that each $\G{}_{(i,r)}$ is a $q$-outerplanar graph with treewidth $O(q)$~\cite{bodlaender1998partial}, which is bounded. Let $\G{}_r = \bigcup_i \G{}_{(i,r)}$. By Theorem~(\ref{thm:twbounded}), we can solve the problem optimally on each $\G{}_r$, $r = 1, ..., q$, in polynomial time. The algorithm then returns the solution with the largest objective over all $r = 1, ..., q$. Using the fact that
$q$ is fixed, one can verify that the overall running time is polynomial in $n$.

\begin{mybox2}
\begin{theorem}
The algorithm gives a factor $(1 - \epsilon)$ approximation on planar graphs for any fixed $\epsilon > 0$.
\end{theorem}
\end{mybox2}

\begin{proof}(Sketch)
Let $q = 2 \cdot \ceil{1/\epsilon}$. We show that the algorithm gives a $1 - 2/q \geq 1 - \epsilon$ approximation. Let $\mc{P}^*$ be an \assignment{} of agents on $\G{}$ that gives the maximum number of integrated agents. Fix an integer $r \in [1~..~ q]$, and let $\mc{W}_r = \{\mc{W}_{(1, r)}, ..., \mc{W}_{(t+1, r)}\}$ be a partition of the vertex set as described above. Let $\mc{P}_{r}$ be an \assignment{} on $\G{}_r$ that is obtained from the proposed algorithm. We now look at the \assignments{} $\mc{P}_r$ and $\plas{}$, restricted to vertices in $\mc{W}_r$. Specifically, let $\mc{P}_{(i,r)}$ and $\mc{P}^*_{(i,r)}$ be the \assignment{} of agents restricted to the subset $W_{(i,r)}$ under $\mc{P}_r$ and $\mc{P}^*$, respectively. Further, let $\texttt{IoA}(\mc{P}_{(i,r)})$ be the number of integrated agents in $\G{}_{(i,r)}$ under $\mc{P}_r$, and $\texttt{IoA}(\mc{P}^*_{(i,r)})$ be the number of 
integrated agents in $\G{}_{(i,r)}$ under $\mc{P}^*$.

\par Define $\Delta_r = \texttt{IoA} (\mc{P}^*) - \sum_{i = 1}^{t+1} \texttt{IoA}(\mc{P}^*_{(i,r)})$.  Integrated vertices that are left uncounted can only exist on the two adjacent layers between each pair of subgraphs $\G{}_{(i,r)}$ and $\G{}_{(i+1,r)}$, $i = 1,...t$. Let $\mc{V}^*$ be the set of integrated vertices under $\plas{}$. We then have, $$\Delta_r \leq \sum_{j = 0}^{t} \left(\mc{V}^* \cap \mc{V}_{j \cdot q + r}\right) + \left(\mc{V}^* \cap \mc{V}_{j \cdot q + r + 1}\right)$$
It follows that $$\min_{r = 1, ..., q}\{ \Delta_r\} \leq \frac{2}{q} \cdot \texttt{IoA} (\mc{P}^*)$$
One can then verify that 
 $\IoAgent{}(\mc{P}_{r^*}) \geq (1 - \frac{2}{q}) \cdot \IoAgent{}(\mc{P}^*)$ where $r^* = \argmin_{r = 1, ..., q}\{ \Delta_r\}$. Lastly, let $\hat{\mc{P}}$ be an \assignment{} returned by the algorithm, $\hat{\mc{P}} = \argmax_{r} \texttt{IoA}(\mc{P}_r)$. It follows that 
 \begin{equation}
     \texttt{IoA}(\hat{\mc{P}}) \geq (1 - \frac{2}{q}) \cdot \texttt{IoA} (\mc{P}^*)
 \end{equation}
This concludes the proof.
\end{proof}

\section{Experimental Evaluation}
We evaluate the empirical performance of the proposed local improvement algorithm for \maxinte{}-\IoAgent{} under several scenarios. Our results demonstrate the high effectiveness of the algorithm on both synthetic and real-world networks.

\subsection{Experimental setup}
\textbf{Networks.} We selected networks based on their sizes and application domain, as shown in Table~(\ref{tab:networks}). Specifically, \texttt{Gnp} and \texttt{Power-law} are synthetic networks generated using the Erdős-R\`{e}nyi~\cite{erdHos1959random} and Barab{\'a}si-Albert~\cite{barabasi1999emergence}
models, respectively. \texttt{City} is a synthetic network of a residential area in Charlottesville, obtained from the Biocomplexity Institute at the University of Virginia; here, vertices are houses, and any pair of houses within $100$ yards are 
considered as neighbors. \texttt{Arena} and \texttt{Google+} are mined social networks obtained from a public repository~\cite{snapnets}.

\noindent
\textbf{Algorithms}. We evaluate the performance of \textsf{Local-Improvement} algorithm using the following baselines: (1) \textsf{Greedy}: Initially, all vertices are occupied by type-\texttt{2} agents; then iteratively $k$ of these are replaced by type-\texttt{1} agents in a greedy manner. Specifically, in each iteration, a replacement that causes the largest increase in the objective value is chosen.  (2) \textsf{Random}:  a random subset of $k$ vertices are chosen for type-\texttt{1} agents, and the remaining vertices are assigned to type-\texttt{2} agents. 

\smallskip

\begin{center}
 \begin{tabular}{||l c c c c||} 
 \hline
 \textbf{Network} & \textbf{Type} & $n$ & $m$ & \textit{Max deg}\\ [0.5ex] 
 \hline 
  
  \texttt{Gnp} & Random & $1,000$ & $4,975$ &  $36$\\ \hline
  
  \texttt{Power-law} & Random & $1,000$ & $5,015$ & $355$\\ \hline

  \texttt{City} & Residential  & $7,444$ & $238,802$ & $165$\\ \hline
  
  \texttt{Arena} & Social & $10,680$ & $24,316$ & $205$ \\ \hline
  
  \texttt{Google+} & Social & $23,613$ & $39,182$ & $2,761$ \\ \hline
\end{tabular}

\captionof{table}{List of networks}
\label{tab:networks}
\end{center}

\noindent
\textbf{Evaluation metrics}. We use two metrics to quantify the performance of algorithms: $(i)$ the {\em integration ratio} $\mu = obj / n$ (i.e., the fraction of integrated agents) and $(ii)$ the \textit{empirical approximation ratio} $\gamma = obj / OPT$ where $OPT$ is the optimal value. The value $OPT$ is computed by solving an integer linear program (ILP) using \textsf{Gurobi}~\cite{Gurobi-2021}.

\noindent
\textbf{Machine and reproducibility.} Experiments were performed on an Intel Xeon(R) Linux machine with 64GB of RAM. The source code and selected datasets are at \url{https://github.com/bridgelessqiu/Integration_Max}.

\subsection{Experimental results}
We present an overview of the results under the following experimental scenarios. 

\noindent
\textbf{Empirical ratio across networks.} We first study the empirical approximation ratio $\gamma$ on different networks. 
For the three large networks, namely \texttt{City},
\texttt{Arena} and \texttt{Google+},
the ILP solver didn't terminate even though it was
run for 24 hours.
Therefore, we restricted our focus to {\em smaller subgraphs}
of these networks. For each subgraph, we fixed the number $k$ of minority agents to be $10\%$ of $n$, where $n$ is the number of vertices in the network. The empirical ratio for each algorithm is then averaged over $100$ repetitions. 

\begin{figure}[!ht]
  \centering
    \includegraphics[width=0.5\textwidth]{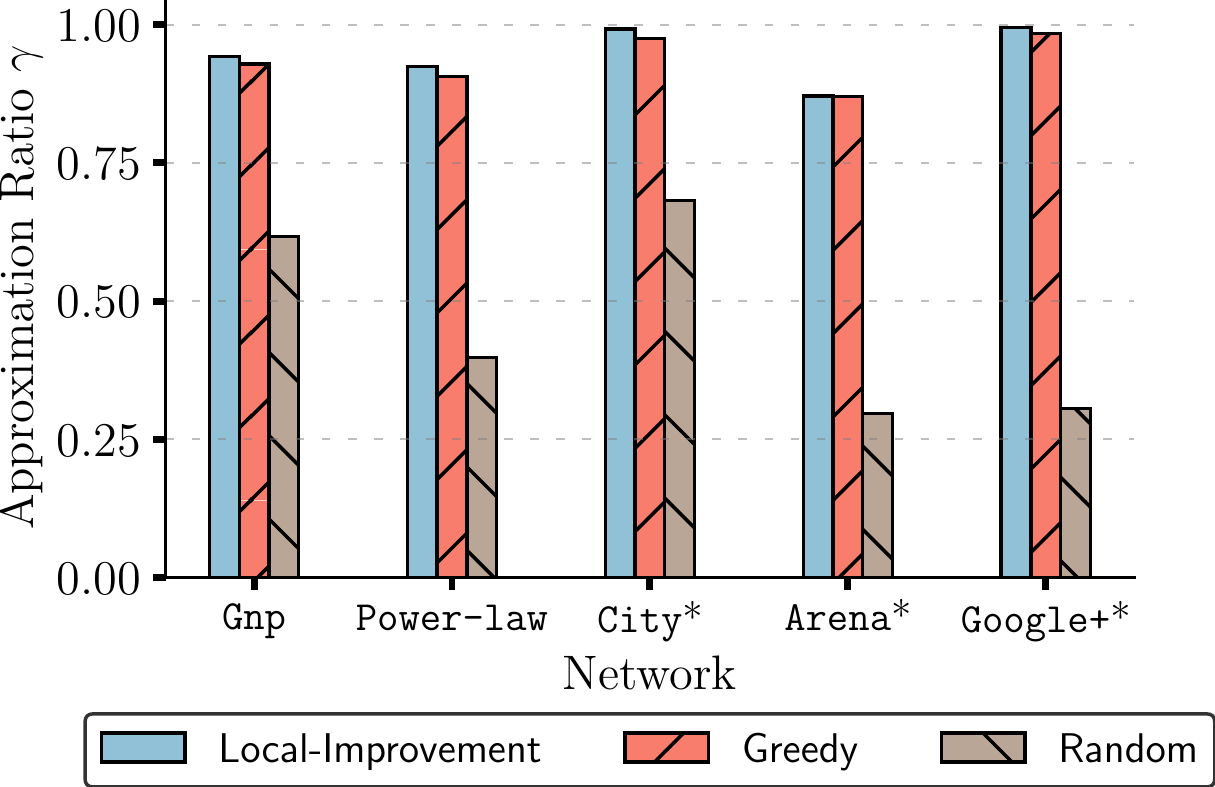}
    \caption{The empirical approximation ratio $\gamma$ for algorithms. The number of vertices and edges ($n$, $m$) for each subgraph are as follows. \texttt{City}*: $(1607, 50112)$, \texttt{Arena}*: $(1981, 9132)$, \texttt{Google+}*: ($2000$, $5042$).}
    \label{fig:empirical-ratio}
\end{figure}

\par Representative results for the empirical ratio  are shown in Fig.~(\ref{fig:empirical-ratio}).
Overall, we observe that the effectiveness of \textsf{Local-Improvement} and \textsf{Greedy} are close to the optimal value, with \textsf{Local-Improvement} outperforming \textsf{Greedy} by a small margin. Specifically, the empirical ratio of \textsf{Local-Improvement} is greater than $0.85$ on all tested instances. 
As one would expect, the empirical ratio of \textsf{Random} is much lower than its counterparts.
Overall, we note that the empirical ratio of \textsf{Local-Improvement} is much higher than its theoretical guarantee of $1/2$. 
Recall from Section~(\ref{sec:general_graphs}) that there are instances where \textsf{Local-Improvement} produces solutions that are of $1/2$ of the optimal value.
Our experimental findings indicate such worst-case instances did not occur in these experiments. We also note that empirically \textsf{Greedy} is comparable to \textsf{Local Improvement}. However, no known performance guarantee for \textsf{Greedy} has been established. In contrast, as shown in Section~(\ref{sec:general_graphs}), \textsf{Local Improvement} provides a guarantee of $1/2$.

\begin{figure}[h!]
\centering
\begin{subfigure}{.38\textwidth}
  \centering
  \includegraphics[width=0.95\linewidth]{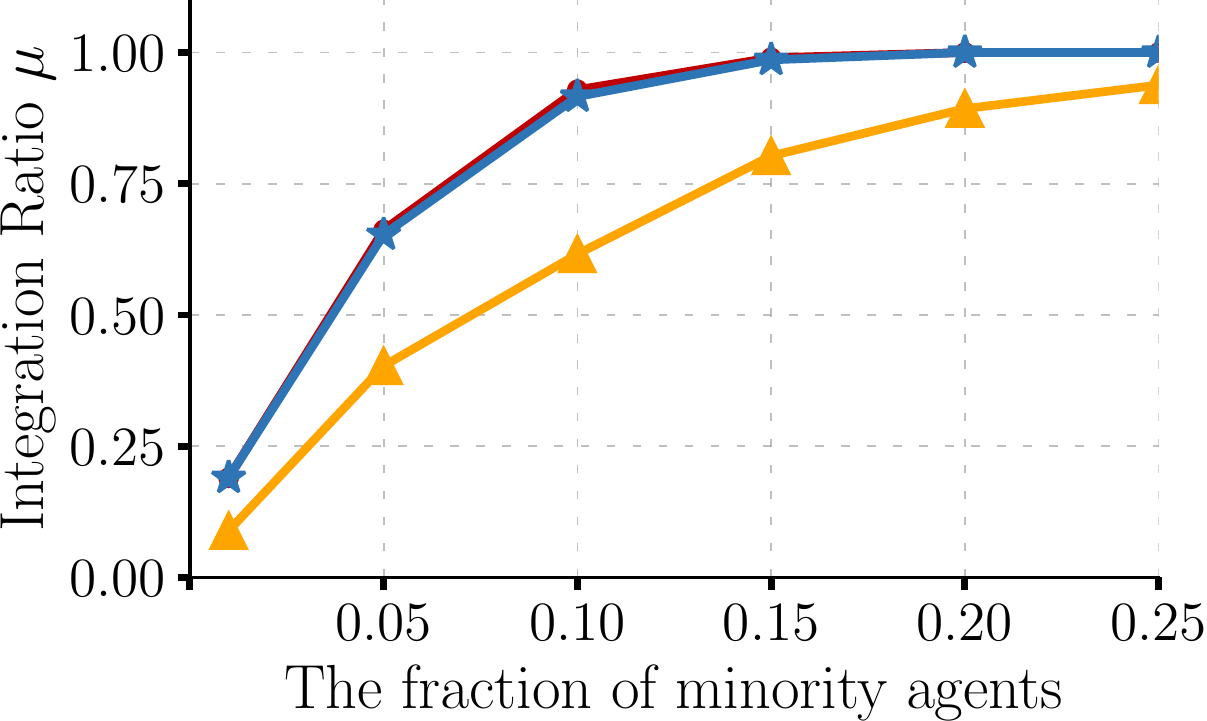}
  \caption{\texttt{Gnp} network}
\end{subfigure}%
\begin{subfigure}{.38\textwidth}
  \centering
  \includegraphics[width=0.95\linewidth]{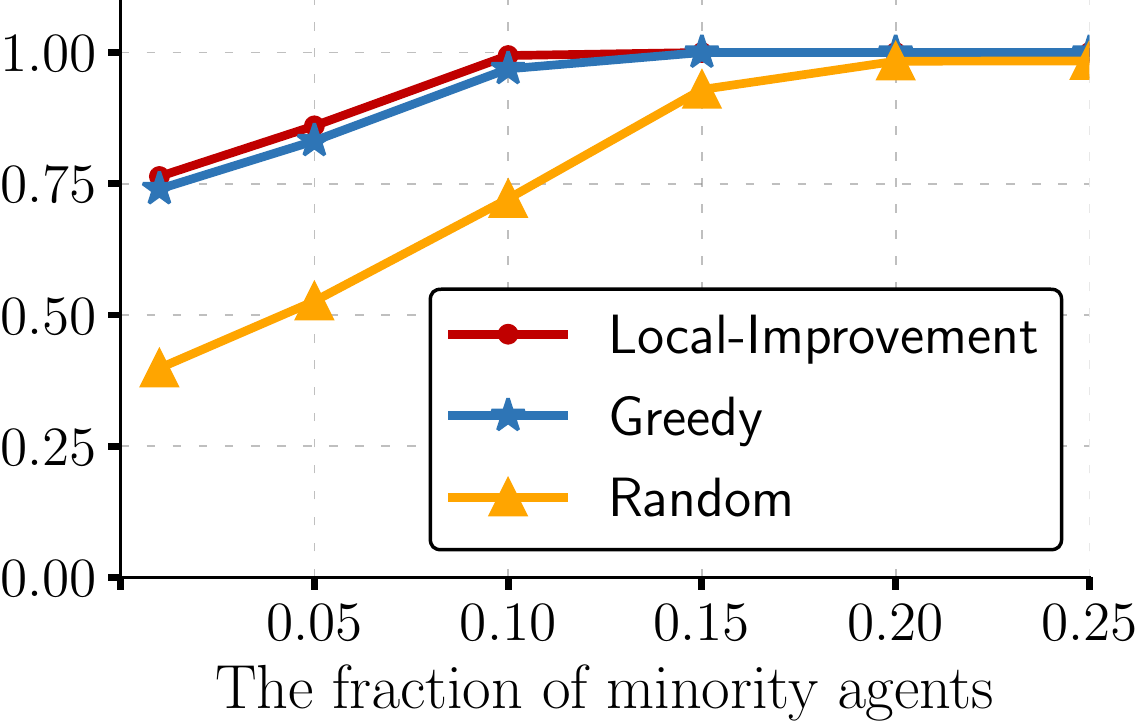}
  \caption{\texttt{City} network}
\end{subfigure}
\caption{The change of the fraction of integrated agents as the fraction of minority agents increases. The networks are \texttt{Gnp} and \texttt{City} shown in Table~(\ref{tab:networks}).}
\label{fig:vary-k}
\end{figure}

\noindent
\textbf{Variations on the number of minority agents.} Next, we study the integration ratio $\mu$ obtained by the algorithms under the scenario where the fraction of minority agents $(k)$ increases from $0.01$ to $0.25$. The representative results for \texttt{Gnp} and \texttt{City} networks are shown in Fig.~(\ref{fig:vary-k}). Overall, we observe that as the fraction of minority agents increases, the integration ratio $\mu$ grows monotonically for all algorithms. Similar results are observed for all the chosen networks. Despite the monotonicity observed in the experiments, we remark that the objective value that an algorithm can obtain is general non-monotone as $k$ increases. 
(A simple example is a star where the objective is maximized for $k = 1$ when the type-\texttt{1} agent is placed at the center. It is easy to verify that as $k$ increases, the optimal objective decreases.) 

\begin{figure}[!ht]
  \centering
    \includegraphics[width=0.5\textwidth]{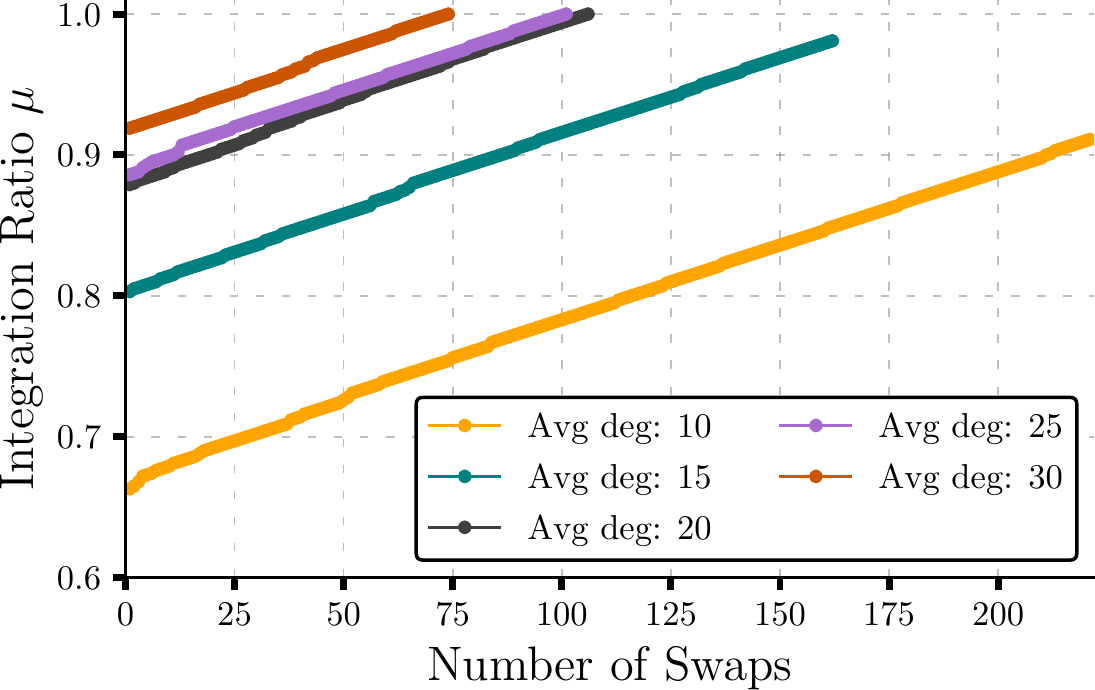}
    \caption{The change in the number of integrated agents as \textsf{Local-Improvement} proceeds. The underlying  \texttt{gnp} networks have $1,000$ vertices; the average degree varies from $10$ to $30$.}
    \label{fig:vary-step}
\end{figure}

\noindent
\textbf{Change of objective as local improvement proceeds}. Lastly, we study the increase in the objective value as the number of swaps used in \textsf{Local-Improvement} is increased.  
Results are shown in Fig.~(\ref{fig:vary-step}) for \texttt{gnp} networks
with 1000 nodes and average degrees varying from 10 to 30.
Overall, we observe a linear relationship between the objective value and the number of swaps.

\section{Conclusions}
\label{sec:concl}
We considered an optimization problem that arises in the context of placing agents on a network to maximize the integration level. 
Since the general problem is \np-hard, we presented approximation algorithms with provable performance guarantees for several versions of the problem.
Our work suggests several directions for further research.
First, it is of interest to investigate approximation algorithms with better performance guarantees for the general problem. One possible approach is to consider local improvement algorithms that instead of swapping just one pair of vertices to increase the number of integrated vertices, swap up to $j$ pairs, for
some fixed $j \geq 2$ in each iteration.
One can also study the problem under network-based
extensions of other integration indices proposed in the social science literature~\cite{Massey-Denton-1988}.
Another direction is the scenario where the total number of agents is less than the number of nodes (so that some nodes remain unoccupied by agents).
In addition, one can also study the  variant where there are agents of three or more types, and the notion of integration is defined by requiring the neighborhood of an agent to include a certain number of agents of the other types. Overall, this  topic offers a variety of interesting new problems for future research.

\bibliographystyle{plain}
\bibliography{bib}

\begin{center}
\fbox{{\Large\textbf{Appendix}}}
\end{center}

\setcounter{section}{3}
\section{Additional Materials for Section 4}

\begin{center}
 \begin{tabular}{||l | l ||} 
 \hline
 \textbf{Notation} & \textbf{Definition} \\ [0.5ex] 
 \hline 
  
  $\mc{P}$ & An assignment return by the algorithm\\ \hline
  $\mc{P}^*$ & An optimal assignment \\ \hline
  $\Gv{}_i(\mc{P})$ & The set of type-$i$ vertices under $\mc{P}$ \\ \hline 
  $\GvU{i}(\mc{P})$ & The set of uncovered type-$i$ vertices under $\mc{P}$ \\ \hline
  $\NU{v}(\mc{P})$ & The set of neighbors of $v \in \Gv{}$ that are uncovered under $\mc{P}$ \\ \hline 
  $\Gamma_v(\mathcal{P})$ & The set of different-type neighbors of $v$ that are uniquely covered by $v$ \\ \hline 
  \hline
  Type-$i$ vertex (under $\mc{P}$) & A vertex occupied by a type-$i$ agent \\ \hline
  An {\em uncovered} vertex (under $\mc{P}$) & A vertex that is not integrated \\ \hline
\end{tabular}

\captionof{table}{A notation table}
\label{tab:notations}
\end{center}

\noindent
Agarwal et al.~\cite{agarwal2020swap} establish that \maxinte{}-\IoAgent{} is \textbf{NP}-hard\footnote{The work by Agarwal et al.~\cite{agarwal2020swap} did not attempt to address the hardness of \maxinte{}-\IoAgent{}, as \IoAgent{} is not the main result in that paper.}. We now further study its solvability. For convenience in presenting the proofs, we define an {\em \assignment{}} from the perspective of vertices of the underlying graph, rather than the perspective of the agents. We remark that the two definitions are equivalent. 


\noindent
\paragraph{Assignment.} An \assignment{} $\mathcal{P} : \Gv{} \rightarrow  \mc{A}$ is a function that assigns an {\em agent type} in $\{1, 2\}$ to each vertex (location) in $\Gv{}$, such that $k$ vertices are assigned type-\texttt{1} and $n-k$ vertices are assigned type-\texttt{2}. 
Given an \assignment{} $\mathcal{P}$, we call a vertex $v$ a {\em type-\texttt{1}} (or {\em type-\texttt{2}}) vertex if $\mathcal{P}(v) = 1$ (or $\mathcal{P}(v) = 2$). Let $\Gv{}_1(\mathcal{P})$ and $\Gv{}_2(\mathcal{P})$ denote the set of type-\texttt{1} and type-\texttt{2} vertices under $\mc{P}$. 
Let $\GvU{1}(\mathcal{P})$ and $\GvU{2}(\mathcal{P})$ denote the set of uncovered type-\texttt{1} and type-\texttt{2} vertices under $\mc{P}$. 
For each vertex $u$, let $\NU{u}(\mathcal{P})$ denote the set of neighbors of $u$ that are uncovered under $\mc{P}$,
and let $\Gamma_u(\mathcal{P})$ denote the set of different-type neighbors of $u$ that are uniquely covered by $u$,
i.e., $\Gamma_u(\mathcal{P})$ is the set of vertices $v$ such that $(i)$ $v$ is a neighbor of $u$, $(ii)$ the type of $v$ is different from the type of $u$, and $(iii)$ $v$ has no other neighbors whose types are the same as $u$'s type. 

\subsection{Analysis of the algorithm} 
\begin{figure}[!h]
  \centering
    \includegraphics[width=0.35\textwidth]{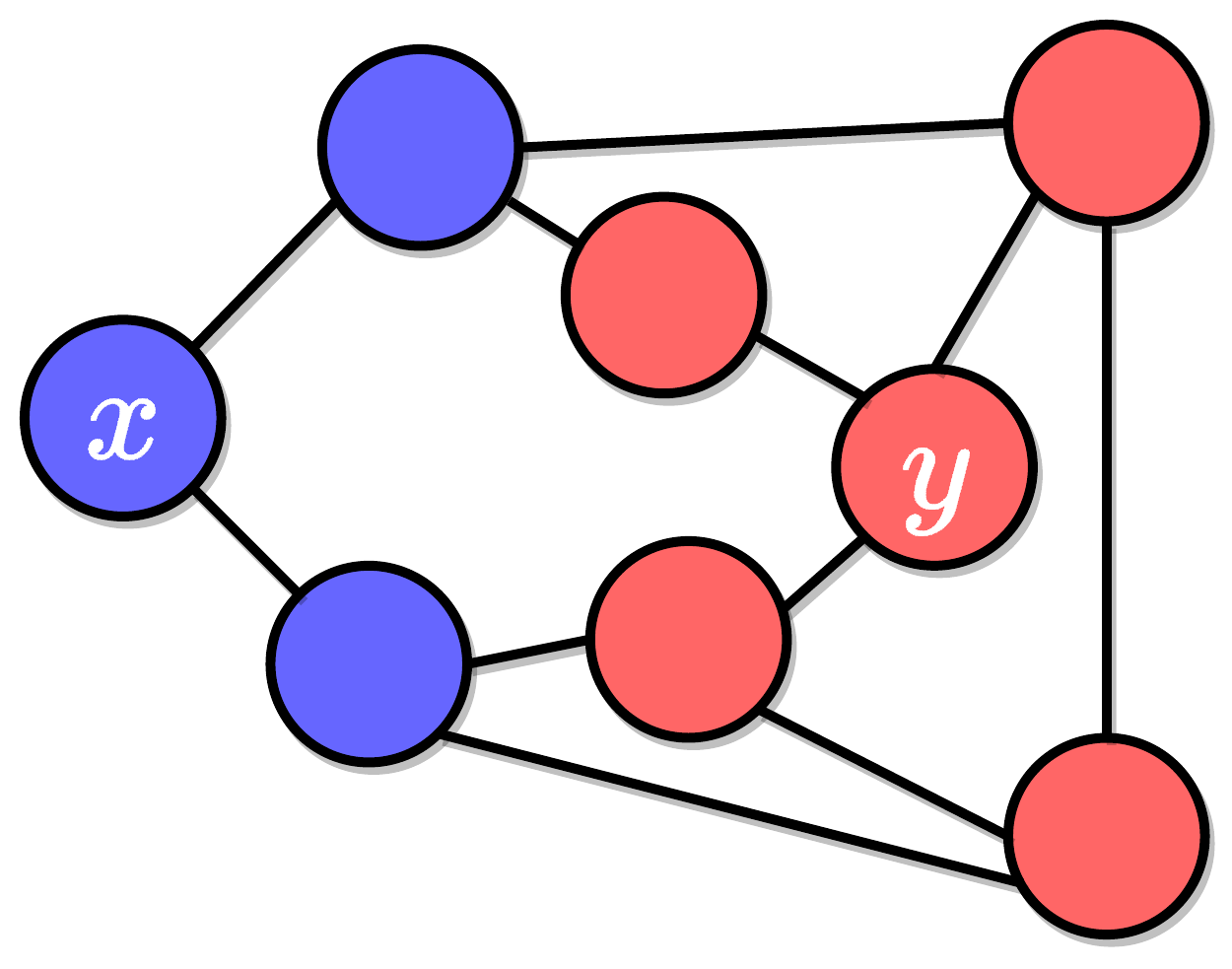}
    \caption{An assignment where type-\texttt{1} and type-\texttt{2} vertices are highlighted in blue and red, respectively. In this example, $\GvU{1}(\mc{P}) = \{x\}$ and $\GvU{2}(\mc{P}) = \{y\}$. Note that this example is {\em only to demonstrate how the two sets are defined}, as later we show that if both $\GvU{1}(\mc{P}) \neq \emptyset$ and $\GvU{2}(\mc{P}) \neq \emptyset$, this \assignment{} cannot be a saturated \assignment{}.}
    \label{fig:example-z}
\end{figure}

We now investigate the performance of Algorithm~(\ref{alg:algo1-IoA}). Let $\mc{P}$ be a saturated \assignment{}\footnote{An \assignment{} is {\em saturated} if no pairwise swap of types between a type-\texttt{1} and a type-\texttt{2} vertices can increase the objective.} returned by Algorithm~(\ref{alg:algo1-IoA}). 
All the analyses are given under $\mc{P}$ unless specified otherwise. 
Recall that $\GvU{1}(\mc{P}) \subseteq \Gv{}$ is the set of type-\texttt{1} vertices are \textit{not} integrated under $\mc{P}$. That is, for each vertex $x \in \GvU{1}(\mc{P})$, all neighbors of $x$ under $\mc{P}$ are also of type-\texttt{1}. Similarly, let $\GvU{2}(\mc{P}) \subseteq \Gv{}$ be the set of type-\texttt{2} vertices who are \textit{not} integrated under $\mc{P}$. An example of such sets are given in Figure~(\ref{fig:example-z}). 

\begin{mybox2}
\begin{observation}
The index $\IoAgent{}(\mc{P}) = n - |\GvU{1}(\mc{P})| - |\GvU{2}(\mc{P})|$.
\end{observation}
\end{mybox2}

\noindent
We now consider the following mutually exclusive and collectively exhaustive cases of $\GvU{1}(\pla{})$ and $\GvU{2}(\pla{})$ under the saturated \assignment{} $\mc{P}$. We start with a simple warm-up case where all the type-\texttt{2} vertices under $\pla{}$ are integrated. 

\noindent
\underline{\textbf{Case 1}}: $\GvU{2}(\pla{}) = \emptyset$. 

\noindent
Under this case, all vertices in $\Gv{}_2(\pla{})$ are integrated which gives 
\begin{equation}
    \IoAgent{}(\mc{P}) \geq |\GvU{2}(\pla{})|  = n - k \geq \frac{1}{2} \cdot n
\end{equation}

\noindent
The above case trivially implies a $2$-approximation of the algorithm. We now consider the case where $\GvU{2}(\pla{}) \neq \emptyset$. 

\bigskip
\noindent
\underline{\textbf{Case 2}}: $\GvU{2}(\pla{}) \neq \emptyset$. 

\noindent
Under this case, there exists at least one vertex in $\Gv{}_2(\pla{})$ that is not integrated. We now study the approximation ratio.

\begin{mybox2}
\begin{lemma}\label{lemma:ioa-1}
For a saturated \assignment{} $\mc{P}$, 
if $\GvU{2}(\pla{}) \neq \emptyset$, then $\GvU{1}(\pla{}) = \emptyset$.
\end{lemma}
\end{mybox2}

\begin{proof}
Let $y \in \GvU{2}(\pla{})$ be a vertex of type-\texttt{2} that is not integrated (i.e., all neighbors of $y$ are of type-\texttt{2}). For contradiction, suppose $\GvU{1}(\pla{}) \neq \emptyset$.. Now let $x \in \GvU{1}(\pla{})$ be an non-integrated vertex of type-\texttt{1} whose neighbors are all of type-\texttt{1}. Let $\mc{P}'$ denote the \assignment{} where we switch the types between $x$ and $y$, that is, $\mc{P}'(x) = \mc{P}(y)= \texttt{2}$, $\mc{P}'(y) = \mc{P}(x) = \texttt{1}$, while the types of all other vertices remain unchanged.

\begin{claim}
$\IoAgent{}(\mc{P}') \geq \IoAgent{}(\mc{P}) + 2$, that is, switching the types of $x$ and $y$ increases the index \IoAgent{} by at least $2$.
\end{claim}

\noindent
We now establish the above claim. First observe that after the switch, only the integration status of vertices in $\{x, y\} \cup \mc{N}(x) \cup \mc{N}(y)$ can change, where $\mc{N}(x)$ and $\mc{N}(y)$ are neighbors of $x$ and $y$. Given that all neighbors of $x$ are of type-\texttt{1} under $\mc{P}$, and $y$ is of type-\texttt{2}, switching $\mc{P}(x)$ with $\mc{P}(y)$ can only increase the number of integrated neighbors in $\mc{N}(x)$. Similarly, switching $\mc{P}(y)$ with $\mc{P}(x)$ can only increase the number of integrated neighbors in $\mc{N}(y)$. Further, note that $x$ (who was not integrated in $\mc{P}$) will be integrated after the switch, as $\mc{N}(x)$ consists of (only) vertices of type-\texttt{1}. By the same argument, $y$ (who was again not integrated in $\mc{P}$) will be integrated after the switch. It follows that after the switch, the index \IoAgent{} would increase by at least $2$, that is, $\IoAgent{}(\mc{P}') \geq \IoAgent{}(\mc{P}) + 2$. This conclude the claim. One may check Figure~(\ref{fig:example-z}) for a visualization.  

\par The claim implies the existence of an improvement move from $\mc{P}$, which contradicts $\mc{P}$ being a saturated \assignment{}. 
It follows that no such an $x \in \GvU{1}(\pla{})$ exists and thus $\GvU{1}(\pla{}) = \emptyset$. 
\end{proof}

\noindent
Lemma~(\ref{lemma:ioa-1}) immediately implies that under case 2 (i.e., $\GvU{2}(\pla{}) \neq \emptyset$), we must have $\GvU{1}(\pla{}) = \emptyset$. 
\begin{equation}
    \IoAgent{}(\mc{P}) \geq |\mc{V}_1| = k
\end{equation}

\noindent
We now argue for a stronger approximation ratio of $2$. Consider the following two mutually exclusive and collectively exhaustive subcases under Case $2$. Recall that for each vertex $x$, $\Gamma_x(\pla{})$ is the set of different-type neighbors of $x$ that are uniquely covered (i.e. ``made integrated'') by $x$ under $\pla{}$. Formally, if $x \in \Gv{}_1(\pla{})$, then $\Gamma_x(\pla{}) = \{y \in \Gv{}_2(\pla{}) \cup \mc{N}(x) \; : \; y \notin \bigcup_{x' \in \Gv{}_1(\pla{}) \setminus \{x\}} \mc{N}(x')\}$

\bigskip
\noindent
\underline{\textbf{Subcase 2.1:}} $\GvU{2}(\pla{}) \neq \emptyset$, and 
$$
    \Gamma_x(\pla{}) \neq \emptyset, \; \; \forall x \in \Gv{}_1(\pla{})
$$
\noindent
that is, for each type-\texttt{1} vertex $x \in \Gv{}_1(\pla{})$, there is at least one type-\texttt{2} neighbors $y$ of $x$ that is uniquely covered (i.e. ``made integrated``) by $x$. 

\noindent
Recall that $\mc{P}$ is a saturated \assignment{} returned by the algorithm. 
By Lemma~(\ref{lemma:ioa-1}), we know that all vertices in $\Gv{}_1(\pla{})$ are integrated under $\mc{P}$. 
Thus, the total number of integrated vertices under $\mc{P}$ equals
$|\Gv{}_1(\pla{})| = k$ plus the number of vertices in $\Gv{}_2(\pla{})$ that are adjacent to vertices in $\Gv{}_1(\pla{})$ (It immediately follows that  $\IoAgent{}(\mc{P}) \geq 2 \cdot \frac{k}{n}$). Let $\mc{P}^*$ by an optimal assignment that gives the maximum number of integrated vertices. We now argue that $\IoAgent{}(\mc{P}) \geq \frac{1}{2} \cdot \IoAgent{}(\mc{P}^*)$.

\par
\begin{figure}[!h]
  \centering
    \includegraphics[width=0.7\textwidth]{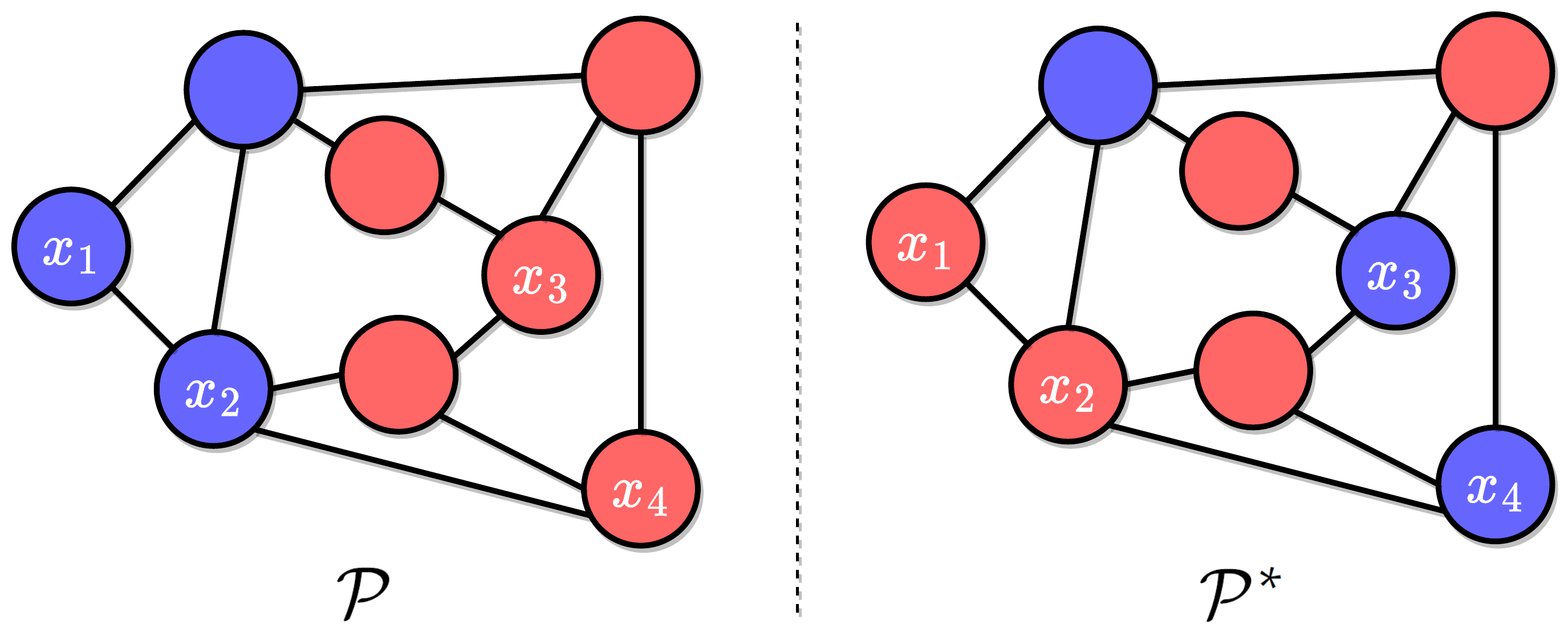}
    \caption{Two assignments $\pla{}$ and $\plas{}$ where type-\texttt{1} and type-\texttt{2} vertices are highlighted in blue and red, respectively. In this case, $\Vto{}= \{x_3, x_4\}$ and $\Vot{} = \{x_1, x_2\}$. We may then transform $\pla{}$ into $\plas{}$ by swapping types between the pair $(x_1, x_3)$ and between $(x_2, x_4)$. Note that this example is {\em only to demonstrate how $\Vto{}$ and $\Vot{}$ are defined}, as  $\pla{}$ cannot be a saturated \assignment{} returned by the algorithm.}
    \label{fig:p-ps}
\end{figure}

 Suppose $\pla{} \neq \plas{}$, that is, for some $x \in \Gv{}$, $\pla{}(x) \neq \plas{}(x)$. 
Let $\Vto{} = \{v \in \Gv{} \; : \; \mc{P}(v) = 2, \mc{P}^*(v)  = 1\}$ be the set of vertices that are type-\texttt{2} under $\pla{}$, but are type-\texttt{1} under $\plas{}$. 
Analogously, let $\Vot{} = \{v \in \Gv{} \; : \; \mc{P}(v) = 1, \mc{P}^*(v) = 2\}$ be the set of vertices of type-\texttt{1} under $\pla{}$, but are of type-\texttt{2} under $\plas{}$. Observe that $|\Vto{}| = |\Vot{}|$.
We may view $\mc{P}^*$ as the result of a transformation from $\pla{}$ under pairwise swaps of types between $\Vto{}$ and $\Vot{}$. An example is given in Figure~(\ref{fig:p-ps}).
We present a key lemma that bounds the difference in the objective value between $\pla{}$ and $\plas{}$.

\begin{mybox2}
\begin{lemma}[Subcase 2.1]\label{lem:subcase1-bound}
Let $\pla{}$ be a saturated \assignment{} that satisfies subcase 2.1, and let $\plas{}$ be an optimal \assignment{}. We have
\begin{align}
    \IoAgent{}(\plas{}) - \IoAgent{}(\pla{})
    &\leq \sum_{y \in \Vto{} \setminus \GvU{2}(\pla{})} |(\mc{N}(y) \cap \GvU{2}(\pla{})| + \sum_{y \in \Vto{} \cap \GvU{2}(\pla{})} \left( |(\mc{N}(y) \cap \GvU{2}(\pla{})| + 1 \right)
\end{align}
\end{lemma}
\end{mybox2}

\begin{proof}
Since $\pla{}$ is saturated, Lemma~(\ref{lemma:ioa-1}) implies that all type-\texttt{1} vertices under $\pla{}$ are integrated.
Thus, $\IoAgent{}(\plas{}) - \IoAgent{}(\pla{})$ is at most the number of type-\texttt{2} vertices that are integrated under $\plas{}$ but are {\em not} integrated under $\pla{}$.

Let $f: \Vot{} \rightarrow  \Vto{}$ be an arbitrary bijective mapping. We may regard $\plas{}$ as a result of the transformation from $\pla{}$ via pairwise swaps of types between vertices specified by $f$ (i.e., the type of $x \in \Vot{}$ is swapped with the type of $f(x) \in \Vto{}$). Observe that only vertices in $\GvU{2}(\pla{})$ that are adjacent to $\Vto{}$ (or within $\Vto{}$) under $\pla{}$ can be newly integrated under $\plas{}$ after swapping $\Vot{}$ with $\Vto{}$ (by the definition of $\GvU{2}(\pla{})$, vertices in $\Vot{}$ have no neighbors in $\GvU{2}(\pla{})$.). It follows that for each vertex $y \in \Vto{}$, at most $|(\mc{N}(y) \cap \GvU{2}(\pla{})|$ of its neighbors can become newly integrated after transforming from $\pla{}$ to $\plas{}$. Further, if also $y \in  \Vto{} \cap \GvU{2}(\pla{})$, $y$ itself could also be newly integrated after the swap. We then have
\begin{align}
    \IoAgent{}(\plas{}) - \IoAgent{}(\pla{}) &\leq |\bigcup_{y \in \Vto{}} \mc{N}(y) \cap \GvU{2}(\pla{})| + |\Vto{} \cap \GvU{2}(\pla{})|\\
    &\leq \sum_{y \in \Vto{} \setminus \GvU{2}(\pla{})} |(\mc{N}(y) \cap \GvU{2}(\pla{})| + \sum_{y \in \Vto{} \cap \GvU{2}(\pla{})} \left( |(\mc{N}(y) \cap \GvU{2}(\pla{})| + 1 \right)
\end{align}

\noindent
where the last inequality follows from the union bound. This completes the proof.
\end{proof}

We note that the bound derived in Lemma~(\ref{lem:subcase1-bound}) is \textit{not} tight for many problem instances. Nevertheless, later we will see that such a bound is enough for our purpose of showing a $\frac{1}{2}$ approximation. Further, we note that there indeed exist a class of problem instances where this bound is exact. 

\noindent
Lemma~(\ref{lem:subcase1-bound}) bounds the maximum difference between $\IoAgent{}(\plas{})$ and $\IoAgent{}(\pla{})$, which is 
$$\sum_{y \in \Vto{} \setminus \GvU{2}(\pla{})} |(\mc{N}(y) \cap \GvU{2}(\pla{})| + \sum_{y \in \Vto{} \cap \GvU{2}(\pla{})} \left( |(\mc{N}(y) \cap \GvU{2}(\pla{})| + 1 \right)$$

\noindent
We now proceed to show that the above difference is {\em at most} $\IoAgent{}(\pla{})$, thereby establishing $\IoAgent{}(\pla{}) \geq \frac{1}{2} \cdot \IoAgent{}(\plas{})$. All the discussion below are under $\pla{}$ unless stated otherwise. Recall that for each vertex $x \in \Gv{}$, $\Gamma_x(\pla{})$ is the set of neighbors of $x$ whose types are different from $x$, and are uniquely covered by $x$ under $\pla{}$. By the definition of Subcase $2.1$, $\Gamma_x(\mathcal{P})$ is not empty for all $x \in \Gv{}_1(\pla{})$. We first argue that for any $y \in \GvU{2}(\pla{})$ and any $x \in \Gv{}_1(\pla{})$, we have $|\mc{N}(y) \cap \GvU{2}(\pla{})| \leq |\Gamma_x(\mathcal{P})|$.

\begin{mybox2}
\begin{lemma}[Subcase 2.1]\label{lemma:subcase2.1-1}
Given a saturated \assignment{} $\pla{}$, for any $y \in \GvU{2}(\pla{})$ and any $x \in \Gv{}_1(\pla{})$, we have 
$$|\mc{N}(y) \cap \GvU{2}(\pla{})| \leq |\Gamma_x(\mathcal{P})|$$
\end{lemma}
\end{mybox2}

\begin{proof}
Given that $y$ is not integrated under $\pla{}$, $x$ and $y$ cannot be adjacent. 
Since $\pla{}$ is a saturated \assignment{}, if the types of $x$ and $y$ are to be swapped,
the number of newly integrated vertices would be at most the number of newly non-integrated vertices. We now examine the integration status of vertices in the closed neighborhood of $x$ and $y$ under $\pla{}$ after such a swap:
\[
\text{For } y \text{ and its neighbors} \left\{\begin{array}{lr}
    \text{All vertices in } \mc{N}(y) \cap \GvU{2}(\pla{})  \text{ become} \textit{ newly integrated}\\
    \text{All vertices in } \mc{N}(y) \cap (\mc{A}_2 \setminus \GvU{2}(\pla{})) \text{ remain integrated}\\
    \text{The vertex } y \text{ itself } \text{becomes {\em newly integrated}}
    \end{array}\right\}
\]

\[
\text{For } x \text{ and its neighbors} \left\{\begin{array}{lr}
    \text{All vertices in } \mc{N}(x) \setminus \Gamma_x(\mathcal{P}) \text{ remain integrated}\\
    \text{Some vertices in } \Gamma_x(\mathcal{P}) \text{ \textit{may} become newly \textbf{non}-integrated}\\
    \text{The vertex } x \text{ itself } \text{ \textit{may} become newly \textbf{non}-integrated}
    \end{array}\right\} 
\]

Overall, the number of vertices that are newly integrated is at least $|\mc{N}(y) \cap \GvU{2}(\pla{})| + 1$, and the number of vertices that are newly non-integrated is at most $|\Gamma_x(\mathcal{P})| + 1$. Since $\mc{P}$ is saturated, it follows that:
\begin{equation}
    |\mc{N}(y) \cap \GvU{2}(\pla{})| \leq |\Gamma_x(\mathcal{P})|
\end{equation}
This concludes the proof.
 \end{proof}

\noindent
We now show that for any $y \in \Gv{}_2(\pla{}) \setminus \GvU{2}(\pla{})$ and any $x \in \Gv{}_1(\pla{})$, we have $|\mc{N}(y) \cap \GvU{2}(\pla{})| \leq |\Gamma_x(\mathcal{P})| + 1$.

\begin{mybox2}
\begin{lemma}[Subcase 2.1]\label{lemma:subcase2.1-2}
Given a saturated \assignment{} $\pla{}$, for any $y \in \Gv{}_2(\pla{}) \setminus \GvU{2}(\pla{})$ and any $x \in \Gv{}_1(\pla{})$, we have 
$$|\mc{N}(y) \cap \GvU{2}(\pla{})| \leq |\Gamma_x(\mathcal{P})| + 1$$
\end{lemma}
\end{mybox2}

\begin{proof}
We partition $\Gv{}_2(\pla{}) \setminus \GvU{2}(\pla{})$ into two subsets $\mc{B}$ and $\mc{C}$, as follows. 
Subset $\mc{B}$ is the set of integrated type-\texttt{2} vertices whose neighbors are all integrated under $\mc{P}$, i.e., 
$$
\mc{B} = \{y \in \Gv{}_2(\pla{}) \setminus \GvU{2}(\pla{}) \; : \; \mc{N}(y) \cap \GvU{2}(\pla{}) = \emptyset\}
$$
\noindent
Subset $\mc{C}$, the complement of $\mc{B}$, is the set of integrated type-\texttt{2} vertices with at least one non-integrated neighbor under $\pla{}$, i.e., 
$\mc{C} = \{y \in \Gv{}_2(\pla{}) \setminus \GvU{2}(\pla{}) \; : \; \mc{N}(y) \cap \GvU{2}(\pla{}) \neq \emptyset\}$. 
The Lemma clearly holds if $y \in \mc{B}$ since then $|\mc{N}(y) \cap \GvU{2}(\pla{})| = 0$. We now present a key claim for the case when $y \in \mc{C}$:

\begin{claim}\label{claim:subcase21}
For all vertices $y \in \mc{C}$, no type-\texttt{1} neighbors of $y$ is uniquely covered by $y$ under $\pla{}$ (i.e., $\Gamma_y(\mathcal{P}) = \emptyset$). 
\end{claim}

\noindent
For contradiction, suppose there exists a type-\texttt{1} neighbors $x \in \mc{N}(y) \cap \Gv{}_1(\pla{})$ of $y$ such that $x$ is not adjacent to any other type-\texttt{2} vertices under $\pla{}$. Then by the definition of subcase 2.1 (i.e., each type-\texttt{1} vertex uniquely covers at least one type-\texttt{2} vertex), $x$ is the only type-\texttt{1} neighbor of $y$. One then can easily verify that exchanging the types between $x$ and $y$ strictly increase the objective of $\pla{}$, contradicting the fact that $\pla{}$ is saturated. This conclude the proof of Claim~(\ref{claim:subcase21}).

\par We continue to assume that $y \in \mc{C}$ and consider an objective non-increasing move from $\pla{}$ where we swap the types between $x$ and $y$. If $y$ is a neighbor of $x$ under $\mc{P}$, then by Claim~(\ref{claim:subcase21}), one can verify that the the maximum loss is $|\Gamma_x(\mathcal{P})|$ and the minimum gain is $|\mc{N}(y) \cap \GvU{2}(\pla{})|$. Thus
\begin{equation}
    |\mc{N}(y) \cap \GvU{2}(\pla{})| \leq |\Gamma_x(\mathcal{P})|
\end{equation}

On the other hand, if $y$ is \textbf{not} a neighbor of $x$ under $\mc{P}$, one can verify that the maximum loss is $|\Gamma_x(\mathcal{P})| + 1$ and the minimum gain is $|\mc{N}(y) \cap \GvU{2}(\pla{})|$. Thus
\begin{equation}
    |\mc{N}(y) \cap \GvU{2}(\pla{})| \leq |\Gamma_x(\mathcal{P})| + 1
\end{equation}
This concludes the proof.
\end{proof}

\noindent
We are now ready to establish $\IoAgent{}(\pla{}) \geq \frac{1}{2} \cdot \IoAgent{}(\plas{})$ under Subcase 2.1. 

\begin{mybox2}
\begin{lemma}[Subcase 2.1]\label{lem:subcase2.1-final}
Suppose $\GvU{2}(\pla{}) \neq \emptyset$ and $\Gamma_x(\pla{}) \neq \emptyset, \forall x \in \Gv{}_1(\pla{})$, we have 
$$\IoAgent{}(\pla{}) \geq \frac{1}{2} \cdot \IoAgent{}(\plas{})$$
where $\plas{}$ is an optimal assignment that gives the maximum objective.
\end{lemma}
\end{mybox2}

\begin{proof}
Note that $\Vto{}$ is a subset of $\Gv{}_2(\pla{})$. Further, 
Observe that $\Gamma_x(\mathcal{P})$ are disjoint for different vertices $x \in \Gv{}_1(\pla{})$. Now, by Lemma~(\ref{lem:subcase1-bound}) to and~(\ref{lemma:subcase2.1-2}), We have
\begin{align}
    &\IoAgent{}(\plas{}) - \IoAgent{}(\pla{}) \nonumber\\ 
    &\leq \sum_{y \in \Vto{} \setminus \GvU{2}(\pla{})} |(\mc{N}(y) \cap \GvU{2}(\pla{})| + \sum_{y \in \Vto{} \cap \GvU{2}(\pla{})} \left( |(\mc{N}(y) \cap \GvU{2}(\pla{})| + 1 \right) \;\;\;\;\;\;\;\; (\text{Lemma}~(\ref{lem:subcase1-bound})) \nonumber\\
    &\leq \sum_{y \in \Vto{} \setminus \GvU{2}(\pla{})} (|\Gamma_{f^{-1}(y)}(\mathcal{P})| + 1) + \sum_{y \in \Vto{} \cap \GvU{2}(\pla{})} (|\Gamma_{f^{-1}(y)}(\mathcal{P})| + 1) \;\;\;\;\;\;\;\; (\text{Lemma}~(\ref{lemma:subcase2.1-1}) \; \&~(\ref{lemma:subcase2.1-2})) \nonumber\\
    &= \left(\sum_{y \in \Vto{}} |\Gamma_{f^{-1}(y)}(\mathcal{P})|\right) +  |\Vto{}| \nonumber\\
    &\leq |\Gv{}_2(\pla{}) \setminus \GvU{2}(\pla{})| + |\Gv{}_1(\pla{})|\label{ineq:subcase2.1}\\
    &\leq \IoAgent{}(\pla{})\nonumber
\end{align}
\noindent
where Inequality~(\ref{ineq:subcase2.1}) follows from $|\Vto{}| = |\Vot{}| \leq |\Gv{}_1(\pla{})|$ and $\left(\sum_{y \in \Vto{}} |\Gamma_{f^{-1}(y)}(\mathcal{P})|\right) \leq |\Gv{}_2(\pla{}) \setminus \GvU{2}(\pla{})|$. This concludes the proof.
\end{proof}

\noindent
We now have shown that if $\GvU{2}(\pla{}) \neq \emptyset$ and $\Gamma_x(\pla{}) \neq \emptyset, \forall x \in \Gv{}_1(\pla{})$, the algorithm gives a 2 approximation. We proceed to the last subcase.

\smallskip

\underline{\textbf{Subcase 2.2:}} $\GvU{2}(\pla{}) \neq \emptyset$, and  
$$
    \Gamma_x(\pla{}) = \emptyset, \;\;\; \exists x \in \Gv{}_1(\pla{})
$$
that is, there exists at least one type-\texttt{1} vertex $x \in \Gv{}_1(\pla{})$ such that for each type-\texttt{2} neighbor $y$ of $x$, $y$ is adjacent to at least one type-\texttt{1} vertex {\em other than $x$}.

\begin{mybox2}
\begin{lemma}[Subcase 2.2]\label{lemma:ioa-2}
Under subcase 2.2, for each non-integrated type-\texttt{2} vertex $y \in \GvU{2}(\pla{})$, all type-\texttt{2} neighbors of $y$ are integrated (i.e., $\mc{N}(y) \cap \GvU{2}(\pla{}) = \emptyset$) under $\pla{}$. 
\end{lemma}
\end{mybox2}

\begin{proof}
Given such a $x \in \Gv{}_1(\pla{})$ defined in Subcase 2.2, for contradiction, suppose there is a non-integrated type-\texttt{2} vertex $y \in \GvU{2}(\pla{})$ s.t. at least one type-\texttt{2} neighbor, denoted by $y' \in \mc{N}(y)$, of $y$ is not integrated under $\mc{P}$ (note that all neighbors of $y$ are of type-\texttt{2} since $y$ is not integrated). Now consider a new \assignment{} $\mc{P}'$ where we switch the types between $x$ and $y$. 

\begin{claim}
We have $\IoAgent{}(\mc{P}') \geq \IoAgent{}(\mc{P}) + 1$, that is, after the switch, the index \IoAgent{} would increase by at least $1$.
\end{claim}

We now establish the claim. Similar to Lemma~(\ref{lemma:ioa-1}), only the integration status of vertices in $\{x, y\} \cup \mc{N}(x) \cup \mc{N}(y)$ can change. We first consider the integration states of vertices in $\mc{N}(x)$. Let $\IoAgent{}(\mc{N}(x), \pla{})$ denote the number of integrated vertices in the neighborhood of $x$ under $\mc{P}$, and let $\Delta \IoAgent{}(\mc{N}(x), \mc{P}) = \IoAgent{}(\mc{N}(x), \pla{}') - \IoAgent{}(\mc{N}(x), \pla{})$ denote the change in the integrated vertices in the neighborhood of $x$ after the switch. 
Let $\mc{N}_{\texttt{1}}(x, \mc{P})$ and $\mc{N}_{\texttt{2}}(x, \mc{P})$ denote the set of type-\texttt{1} and type-\texttt{2} neighbors of $x$ under $\mc{P}$, respectively. 

\par Under Subcase $2.2$, each vertex $\mc{N}_{\texttt{2}}(x, \mc{P})$ is adjacent to at least one type-\texttt{1} vertex in additional to $x$, thus, all vertices in $N_{\texttt{2}}(x, \mc{P})$ remain integrated after we swap types between $y$ and $x$. Further, it is easy to see that the swap cannot decrease the number of integrated vertices in $\mc{N}_{\texttt{1}}(x, \mc{P})$. It follows that 
$$
    \Delta \IoAgent{}(\mc{N}(x), \mc{P}) \geq 0
$$
Now consider the integration states of vertices in $\mc{N}(y)$. First observe that $\mc{N}_{\texttt{1}}(y, \mc{P}) = \emptyset$ since $y \in \GvU{2}(\pla{})$ is not integrated. Also, swapping the types between $y$ and $x$ will not decrease the number of integrated vertices in $\mc{N}_{\texttt{2}}(y, \mc{P})$. In fact, since there exists a vertex $y' \in \mc{N}_{\texttt{2}}(y, \mc{P})$ who is not integrated under $\mc{P}$, the swap makes $y'$ integrated (as $x$ and $y'$ are of different types). It follows that 
$$
\Delta \IoAgent{}(\mc{N}(y), \mc{P}) \vcentcolon =  \IoAgent{}(\mc{N}(y), \mc{P}') - \IoAgent{}(\mc{N}(y), \mc{P}) \geq 1
$$

Lastly, we consider the integration states of $x$ and $y$. In particular, $x$ is integrated under $\mc{P}$, and after the swap, it might become non-integrated. On the other hand, $y$ is not integrated under $\mc{P}$, it must become newly integrated after the swap. Nevertheless, the net increase of the number of integrated vertices in $\{x, y\}$ is at least $0$ when we change from $\mc{P}$ to $\mc{P}'$. 
Overall, it follows that
\begin{equation}
    \IoAgent{}(\mc{P}') - \IoAgent{}(\mc{P}) = \Delta \IoAgent{}(\mc{N}(x), \mc{P}) + \Delta \IoAgent{}(\mc{N}(y), \mc{P}) + \Delta \IoAgent{}(\{x, y\}, \mc{P}) \geq 1
\end{equation}

This concludes the claim. Note that the claim implies the existence of an improvement move from $\mc{P}$, which contradicts $\mc{P}$ being a saturated \assignment{} returned by Algorithm~(\ref{alg:algo1-IoA}). Thus, no such a non-integrated type-\texttt{2} vertex $y'$ of $y$ can exist, that is, for each $y \in \GvU{2}(\pla{})$, all (type-\texttt{2}) neighbors of $y$ are integrated. This concludes the proof.
\end{proof}

\noindent
Lemma~(\ref{lemma:ioa-2}) implies that under Subcase 2.2, the vertices in $\GvU{2}(\pla{})$ form an independent set of $\mc{G}$, as stated in the corollary below.

\begin{mybox2}
\begin{corollary}[Subcase 2.2]\label{coro:indpendent-set}
Under Subcase 2.2, vertices in $\GvU{2}(\pla{})$ form an independent set of $\mc{G}$.
\end{corollary}
\end{mybox2}

\noindent
Observe that $\IoAgent{}(\mc{P}) = (n - |\GvU{2}(\pla{})|)$. Using Lemma~(\ref{lemma:ioa-2}), we now argue the size of $\GvU{2}(\pla{})$ cannot be too large.

\begin{mybox2}
\begin{lemma}[Subcase 2.2]\label{lemma:ioa-3}
Under Subcase 2.2, we have
$$|\GvU{2}(\pla{})| \leq \frac{n}{2}$$
\end{lemma}
\end{mybox2}

\begin{proof}
Let 
$$\mc{Y} \vcentcolon= \{y \in \Gv{}_2(\pla{}) \setminus \GvU{2}(\pla{}) \; : \; \mc{N}(y) \cap \GvU{2}(\pla{}) \neq \emptyset\}
$$
be the set of type-\texttt{2} {\em integrated} vertices whose has at least one non-integrated type-\texttt{2} neighbor. Recall that $\Gamma_y(\mathcal{P})$ is the set of type-\texttt{1} neighbors of $y$ who are uniquely covered by $y$ under $\pla{}$. We first note that $\Gamma_y(\mathcal{P})$ (if not empty) are mutually disjoint for different $y \in \mc{Y}$. It follows that 
\begin{equation}
    \IoAgent{}(\mc{P}) \geq |\mc{Y}| + \sum_{y \in \mc{Y}} |\Gamma_y(\mathcal{P})|
\end{equation}
Now revisit the definition of subcase 2.2. In particular there exists a type-\texttt{1} vertex $x \in \Gv{}_1(\pla{})$ such that each type-\texttt{2} neighbor of $x$ is also covered by (i.e., adjacent to) at least one other type-\texttt{1} vertex. Suppose we switch the types between such a vertex $x$ and a vertex $y \in \mc{Y}$, and let $\mc{P}'$ denote the resulting new \assignment{}. Observe the following in $\mc{P}'$

\begin{claim}
All vertices in $\mc{N}(x)$ remains integrated in $\mc{P}'$.
\end{claim}
This holds since these neighbors are either of $(i)$ type-\texttt{1} which are now adjacent to $x$ of type-\texttt{2} in $\mc{P}'$, or of $(ii)$ type-\texttt{2} which are adjacent to at least one other type-\texttt{1} vertex.

\begin{claim}
All vertices in $\mc{N}(y) \cap \GvU{2}(\pla{})$ become newly integrated in $\mc{P}'$, and all vertices in $\Gamma_y(\mathcal{P})$ may become newly non-integrated in $\mc{P}'$. The integration status of all other vertices in $\mc{N}(y)$ remain unchanged from $\mc{P}$ to $\mc{P}'$.
\end{claim}

One can easily verify the above claim based on the fact that $y$ is of type-\texttt{1} under $\pla{}'$. Lastly, note that in $y$ remains integrated in $\mc{P}'$ since $y$ is of type-\texttt{1} in $\mc{P}'$ and has at least one type-\texttt{2} neighbor. On the other hand, $x$ (who was integrated in $\pla{}$) might not be integrated in $\mc{P}'$. It follows that the maximum loss of objective after the swap is $|\Gamma_y(\mathcal{P})| + 1$, where as the minimum gain is $|\mc{N}(y) \cap \GvU{2}(\pla{})|$. Since $\mc{P}$ is a saturated \assignment{} returned by the algorithm, we must have $\texttt{IoA}(\mc{P}) \geq \texttt{IoA}(\mc{P}')$. It follows that 
\begin{equation}\label{eq:1/2-subcase2}
    |\mc{N}(y) \cap \GvU{2}(\pla{})| \leq |\Gamma_y(\mathcal{P})| + 1, \; \forall y \in \mc{Y}
\end{equation}
Lastly, by Corollary~(\ref{coro:indpendent-set}), vertices in $\GvU{2}(\pla{})$ form an independent set of $G$. Thus, 
\begin{equation}
    |\GvU{2}(\pla{})| = |\bigcup_{y \in \mc{Y}} \mc{N}(y) \cap \GvU{2}(\pla{})|
\end{equation}
Overall, we have that
\begin{align}
    |\GvU{2}(\pla{})| &= |\bigcup_{y \in \mc{Y}} \mc{N}(y) \cap \GvU{2}(\pla{})| \;\;\;\;\;\;\;\;\;\;\;\;\;\;\;\;\;\;\;\;\; (\text{Corollary~(\ref{coro:indpendent-set})}\\
    &\leq \sum_{y \in \mc{Y}} |\mc{N}(y) \cap \GvU{2}(\pla{})| \;\;\;\;\;\;\;\;\;\;\;\;\;\;\;\;\;\;\;\;\;\; (\text{Union bound}) \\ 
    &\leq \sum_{y \in \mc{Y}} (|\Gamma_y(\mathcal{P})| + 1) \;\;\;\;\;\;\;\;\;\;\;\;\;\;\;\;\;\;\;\;\; \;\;\;\;\;\; (\text{Eq.~(\ref{eq:1/2-subcase2}}))\\
    &\leq |\Gv{}_1(\pla{})| + |\mc{Y}| \;\;\;\;\;\;\;\;\;\;\;\;\;\;\;\;\;\;\;\;\;\;\; \;\;\;\;\;\;\;\;\;\; (\text{$\Gamma_y(\mathcal{P})$ are mutually disjoint})\\
    &\leq |\Gv{}_1(\pla{})| + |\Gv{}_2(\pla{}) \setminus \GvU{2}(\pla{})| \;\;\;\;\;\;\;\;\; (\text{By } \mc{Y} \subseteq \Gv{}_2(\pla{}) \setminus \GvU{2}(\pla{}))\\
    &= n - |\GvU{2}(\pla{})|
\end{align}
It immediately follows that
\begin{equation}
    |\GvU{2}(\pla{})| \leq \frac{n}{2} 
\end{equation}
This concludes the proof. 
\end{proof}

\noindent
Lastly, Since $\IoAgent{}(\mc{P}) = n - |\GvU{2}(\pla{})|$, by Lemma~(\ref{lemma:ioa-3}), we have $\IoAgent{}(\mc{P}) = n - |\GvU{2}(\pla{})| \geq \frac{1}{2} \cdot n \geq \frac{1}{2} \cdot \IoAgent{}(\plas{})$, thereby establishing a $1/2$ approximation for Subcase 2.2. Overall, we have shown that a saturated \assignment{} $\pla{}$ returned by Algorithm~(\ref{alg:algo1-IoA}) gives a $2$-approximation for \maxinte{}-\IoAgent{}. The Theorem immediately follows.

\begin{mybox2}
\begin{theorem}\label{thm-1/2main}
Algorithm~(\ref{alg:algo1-IoA}) gives a $\frac{1}{2}$-approximation for \maxinte{}-\IoAgent{}.
\end{theorem}
\end{mybox2}


\subsubsection{Analysis is tight} 
We now present a class of problem instances where the approximation ratio of the solution produced by Algorithm~(\ref{alg:algo1-IoA}) can be arbitrarily close to $1/2$. Therefore, the ratio $1/2$ in the statement of Theorem~(\ref{thm-1/2main}) cannot be improved,
so {\em our analysis is tight.}

\begin{figure}[!h]
  \centering
    \includegraphics[width=0.3\textwidth]{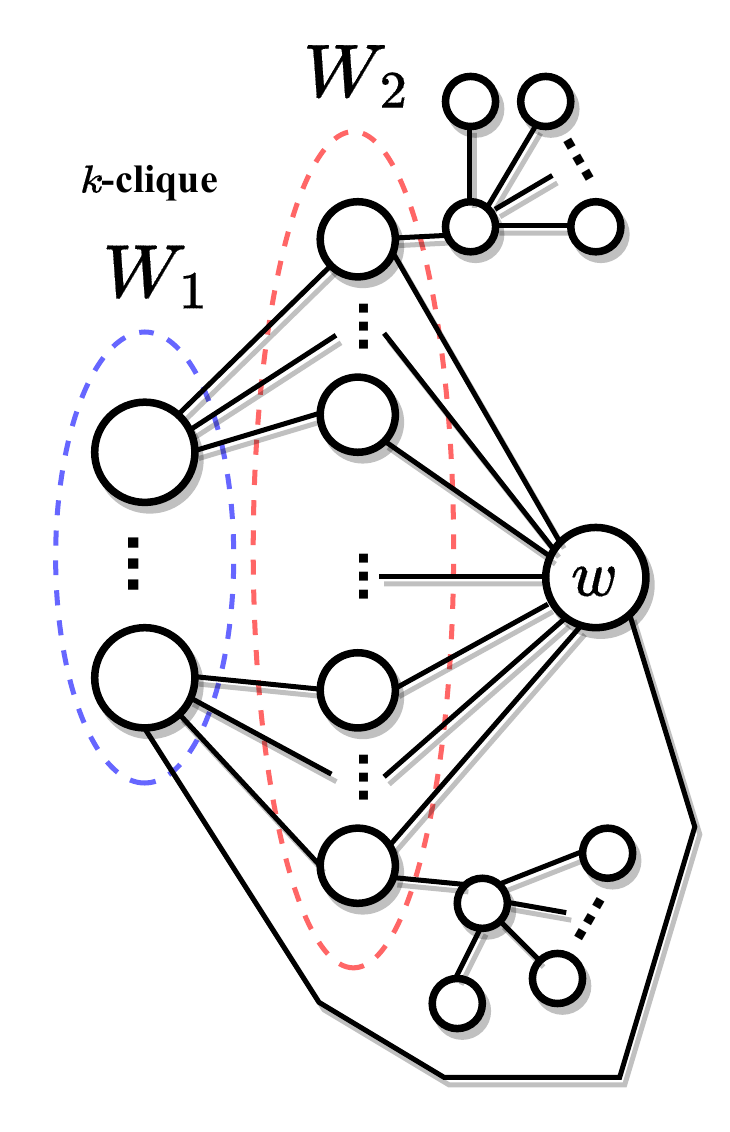}
    \caption{A pictorial example of a problem instance where Algorithm~(\ref{alg:algo1-IoA}) gives an \assignment{} whose approximation ratio is $\frac{1}{2}$.}
    \label{fig:tight}
\end{figure}

\begin{mybox2}
\begin{proposition}
For every $\epsilon > 0$, there exists a problem instance of \maxinte{}-\IoAgent{} for which there is a saturated \assignment{} $\pla{}$ such that $\IoAgent{}(\pla{}) \leq (\frac{1}{2} + \epsilon) \cdot \textsc{OPT}$.
\end{proposition}
\end{mybox2}

\begin{proof}
Recall that $k$ is the number of type-\texttt{1} vertices. We first present the construction of the graph $\G{} = (\Gv{}, \Ge{})$. Let $\W{}_1$ be a set of $k$ vertices that form a clique. 
For each $v \in \W{}_1$, we introduce a set $\mc{U}_v$ of $k$ vertices outside the clique that are adjacent $v$. Let $\W{}_2 = \bigcup_{v \in \W{}_1} \mc{U}_v$ denote the union of these sets. All vertices in $\W{}_2$ are also adjacent to a new vertex $w$, and we further make this vertex $w$ adjacent to exactly one vertex in $\W{}_1$.
Lastly, we added a total of $k$ stars, each of which consists of $k - 1$ vertices (i.e., each star has a center vertex and $k - 2$ leaf vertices). 
We then connect the center of each star to one vertex in a unique $\mc{U}_v$. This completes the construction. An example is given in Figure~(\ref{fig:tight}). 

Now consider an \assignment{} $\pla{}$ where all vertices in $\W{}_1$ are of type-\texttt{1} (recall that $k = |\W{}_1|$), and the rest of vertices are of type-\texttt{2}. One can verify that such an assignment is {\em saturated} (and thus could be returned by Algorithm~(\ref{alg:algo1-IoA})). On the other hand, an \assignment{} $\plas{}$ that gives a strictly higher objective is where we assign $(i)$ type-\texttt{1} to one vertex in $\W{}_1$, $(i)$ assign $w$ to type-\texttt{1}, and $(iii)$ the centers of $k - 2$ stars (with any two stars being left out) are assigned with type-\texttt{1}. The rest of vertices are of type-\texttt{2}.

One can verify that $\IoAgent{}(\pla{}) = k^2 + k + 1$, and $\IoAgent{}(\plas{}) = 2k^2 - 2k + 4$. 
The ratio $\IoAgent{}(\pla{}) / \IoAgent{}(\plas{}) = 1/2$ as $k$ goes to infinity. Since $\IoAgent{}(\plas{}) \leq \text{OPT}$ where \text{OPT} is the optimal objective of a problem instance, the claim follows.
\end{proof}

\section{Additional Material for Section 5}
We study the problem instances when the number of type-\texttt{1} agents is a constant fraction of the total number of agents, that is, $k = \alpha \cdot n$ for some constant $ 0 \leq \alpha \leq 1/2$. We refer to this problem as $\alpha n$-\maxinte{}-\IoAgent{}. For example, $\alpha = 1/2$ implies the {\em bisection} constraint. 

\subsection{Intractability remains}
We first show that $\alpha n$-\maxinte{}-\IoAgent{} problem 
remains intractable.

\begin{mybox2}
\begin{theorem}
The problem $\alpha n$-\maxinte{}-\IoAgent{} is \textbf{NP}-hard.
\end{theorem}
\end{mybox2}

\begin{figure}[!h]
  \centering
    \includegraphics[width=0.4\textwidth]{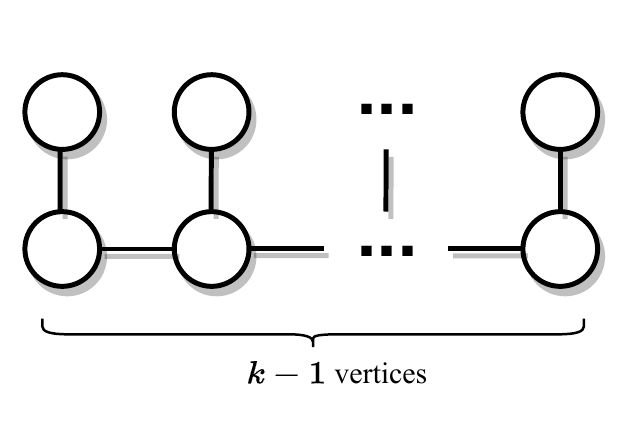}
    \caption{An example of graph $\G{}_2$}
    \label{fig:g1}
\end{figure}

\begin{proof}
We present a reduction from the general \maxinte{}-\IoAgent{} problem to $\alpha n$-\maxinte{}-\IoAgent{} where $\alpha = \frac{1}{2}$. Let $\Pi_1 = \langle \G{}, \A{}, n \rangle$ be an instance of \maxinte{}-\IoAgent{}, $\A{} = \{\A{}_1, \A{}_2\}$, where $k = |\A{}_1|$ is the number of type-\texttt{1} agents that needs to be assigned, and $n = |\Gv{}(\G{})|$ is the total number of agents. The decision question asks whether there exists an \assignment{} of agent-types for $\Pi_1$ such that all the $n$ vertices are integrated. This question is known to be \textbf{NP}-hard~\cite{agarwal2020swap}.

An instance, $\Pi_2 = \langle \G{}', \A{}', 2n \rangle$, $\A{}' = \{\A{}_1', \A{}_2'\}$, of the bisection version of \maxinte{}-\IoAgent{} consists of the following components. To from the graph $\G{}'$, the first component $\G{}_1$ a copy of $\G{}$. Let $\G{}_2$ be a graph formed by first creating a simple path $I$ with $k - 1$ vertices, then for each vertex $v$ on the path, we introduce a new vertex (not on the path) that is uniquely adjacent to $v$. Overall, $\G{}_2$ has $2k - 2$ vertices. An example of $\G{}_2$ is shown in Figure~(\ref{fig:g1}). Let $\G{}_3$ be a star graph with $n - 2k + 2$ vertices (i.e., one center with $n - 2k + 1$ leaf vertices). Lastly, the final graph $\G{}'$ consists of the three aforementioned connected components: $\G{}_1$, $\G{}_2$, and $\G{}_3$. One can verify that $\G{}'$ has $2n$ vertices (and thus the number of agents is $2n$). We set the number of type-\texttt{1} agents $|\A{}_1'| = n$, corresponding to the bisection constraint $|\A{}_1'| = 1/2 \cdot |\A{}'|$.

\par We now argue that $\Pi_1$ admits an \assignment{} where all $n$ vertices are integrated if and only if $\Pi_2$ has an \assignment{} where all $2n$ vertices are integrated. 

\begin{itemize}
    \item[$(\Rightarrow)$] Suppose $\Pi_1$ has an \assignment{} $\mc{P}$ on $\G{}$ such that all vertices are integrated. We now present an \assignment{} $\mc{P}'$ for $\G{}'$ such that all vertices are integrated in $\Pi_2$. Specifically, we discuss how types are assigned on $\G{}_1$, $\G{}_2$, and $\G{}_3$. The \assignment{} of agent-type on $\G{}_1$ is the same as that of on $\G{}$ under $\pla{}$. Next, for $\G{}_2$, we set all the $k-1$ vertices on the path $I$ to type-\texttt{1} (i.e., taken by type-\texttt{1} agents), and the rest of $k-1$ vertices are of type-\texttt{2}. Lastly, for $\G{}_3$, all the $n - 2k + 1$ leaf vertices are of type-\texttt{1}, and the center vertex is of type-\texttt{2}. The completes the construction of $\mc{P}'$. One can verify that the total number of type-\texttt{1} vertices is $k + (k-1) + (n - 2k + 1) = n$, and further, all the $2n$ vertices are integrated under $\mc{P}'$.
    
    \item[$(\Leftarrow)$] Suppose $\Pi_2$ has an \assignment{} $\mc{P}'$ on $\G{}'$ such that all vertices are integrated. We show that there exists an \assignment{} $\mc{P}$ on $\G{}$ such that all vertices are integrated in $\Pi_1$. Consider the \assignment{} $\mc{P}'$ restricted to $\G{}_2$ and $\G{}_3$. We first observe that any \assignment{} that makes all vertices in $\G{}_2$ integrated must has exactly $k - 1$ type-\texttt{1} vertices in $\G{}_2$. As for $\G{}_3$, there are two possible \assignments{} that makes all vertices integrated: either $(i)$ having one type-\texttt{2} vertex at the center and the leaf vertices are of type-\texttt{1} or $(ii)$ vise versa. Note that under the first \assignment{}, the total number of type-\texttt{1} vertices placed on $\G{}_2$ and $\G{}_3$ is $(k - 1) + (n - 2k + 1) = n - k$. Since the total number of type-\texttt{1} vertices is $n$, there are exactly $k$ type-\texttt{1} vertices in $\G{}_1$. Thus, the \assignment{} $\mc{P}$ is obtained by restricting $\mc{P}'$ to $\G{}_1$. On the other hand, under the second type of \assignment{} in $\G{}_3$, the total number of type-\texttt{1} vertices placed in $\G{}_2$ and $\G{}_3$ is $(k - 1) + 1 = k$. That is, there are $n - k$ type-\texttt{1} vertices in $\G{}_1$ under $\mc{P}'$. Then $\mc{P}$ is obtained by flipping the types of vertices (i.e., type-\texttt{1} changes to type-\texttt{2}, vise versa) assigned in $\G{}_1$ under $\pla{}'$. 
\end{itemize}
This concludes the proof.
\end{proof}

\subsection{A semidefinite programming approach}

Our approximation results are given in terms of the {\em expected approximation ratio} $\gamma$. One can obtain a w.h.p. bound by $(i)$ running the algorithm $K$ rounds, and $(ii)$ output the best solution. In particular, one can verify that for each round, the probability of producing an approximation factor $(1-\epsilon) \cdot \gamma$ is at least $1 - (1-\gamma)/(1-(1-\epsilon) \cdot \gamma)$ for arbitrarily small constant $\epsilon > 0$. One can then choose a large enough $K$ to obtain a high probability bound, while $K$ remains a polynomial of $n$.

We now present an approximation algorithm based on a semidefinite programming (SDP) relaxation. Given a graph $\G{} = (\Gv{}, \Ge{})$, each vertex $i \in \Gv{}$ has a binary variable $x_i \in \{-1, 1\}$ such that $x_i = -1$ if $i$ is of type-\texttt{1}, and $x_i = 1$ if $i$ is of type-\texttt{2}. To start with, a quadratic program (QP) of $\alpha n$-\maxinte{}-\IoAgent{} and its SDP relaxation can be formulated as follows:

\noindent
\begin{minipage}{.5\linewidth}
\begin{align*}
\text{QP}: \;\;\;\;\; \text{maximize} \quad
&\sum_{i \in \Gv{}} \; \max_{j \in \mc{N}(i)} \; \{ \frac{1 - x_i x_j}{2} \} \\
\text{s.t.} \quad
&\sum_{i < j} x_i x_j = \frac{(1-2\alpha)^2 \cdot n^2 - n}{2} \\
&x_i \in \{-1, 1\}, \;\;\;\;\; \forall i \in V
\end{align*}
\end{minipage}%
\begin{minipage}{.5\linewidth}
\begin{align*}
\text{SDP}: \;\;\;\;\; \text{maximize} \quad
&\sum_{i \in \Gv{}} \; \max_{j \in \mc{N}(i)} \; \{ \frac{1 - \vec{y}_i \cdot \vec{y}_j}{2} \} \\
\text{s.t.} \quad
& \sum_{i < j} \vec{y}_i \cdot \vec{y}_j \leq \frac{(1-2\alpha)^2 \cdot n^2 - n}{2}\\
& \vec{y}_i \cdot \vec{y}_i = 1, \;\;\;\;\; \forall i \in V
\end{align*}
\end{minipage}

\begin{mybox2}
\begin{observation}
QP is a valid program for the $\alpha n$-\maxinte{}-\IoAgent{} problem. Further, if $\textsc{OPT}_{QP}$ and $\textsc{OPT}_{SDP}$ are the optimal solutions to QP and SDP, respectively, we have $\textsc{OPT}_{SDP} \geq \textsc{OPT}_{QP}$.
\end{observation}
\end{mybox2}

Note that a naive constraint for an $(\alpha n, (1-\alpha) n)$-paritition is $\sum_{i} x_i = (1 - 2\alpha) \cdot n$. With a simple derivation, one can verify that the constraint $\sum_{i < j} \vec{y}_i \cdot \vec{y}_j \leq \frac{(1-2\alpha)^2 \cdot n^2 - n}{2}$ is equivalent to the constraint $\sum_{i} x_i = (1- 2\alpha) \cdot n$, as follows:
\begin{equation}
    \left( \sum_i x_i \right)^2 = \sum_{i} (x_i)^2 + 2 \sum_{i < j} x_i x_j = n +  2 \sum_{i < j} x_i x_j = (1 - 2 \alpha)^2 \cdot n^2
\end{equation}

\noindent
To see that the SDP formulation is indeed a relaxation of the QP, given any feasible solution of the QP, we can construct a feasible solution of the SDP as follows. For each $x_i$, we set the first entry in the corresponding $\vec{y}_i$ to equal the value of $x_i$, and the remaining entries in $\vec{y}_i$ to $0$. One can verify that the two solutions have the same objective. It follows that for each solution of the QP, there is a corresponding solution of the SDP with the same objective, thus, $\textsc{OPT}_{SDP} \geq \textsc{OPT}_{QP}$. 


\subsubsection*{First step: Rounding the SDP}
We first solve the proposed SDP and obtain the set of vectors $\{\vec{y}_1, ..., \vec{y}_n\}$. Let $\textsc{OPT}_{SDP}$ be the optimal objective of the SDP. We want a partition $\{\Gv{}_1, \Gv{}_2\}$ of the vertex set such that vertices in $\Gv{}_i$ are of type-$i$, $i = \texttt{1}, \texttt{2}$. To do so, we apply Goemans and Williamson's {\em hyperplane rounding method}~\cite{goemans1995improved}. In particular, we draw a a random hyperplane thought the origin with a normal vector $r$, and then $\Gv{}_1 = \{i : \vec{y}_i \cdot r \geq 0\}$ and $\Gv{}_2 = \{i : \vec{y}_i \cdot r < 0\}$. This rounding method has the following desirable property:

\begin{mybox2}
\begin{lemma}[Goemans and Williamson~\cite{goemans1995improved}] \label{lemma:GW}
The probability that two vertices $i$ and $j$ being in different subsets is $1/\pi \cdot \arccos{} (\vec{y}_i \cdot \vec{y}_j)$.
\end{lemma}
\end{mybox2}
\vspace{-10px}

Consider an \assignment{} $\pla{}$ generated by the above rounding method (i.e., vertices in $\Gv{}_i$ are assigned to type-$i$). Let $f(\Gv{}_1) : 2^{\Gv{}} \rightarrow \mathbb{N}$ be the number of integrated vertices under such an \assignment{}. Based on Lemma~(\ref{lemma:GW}), we argue that $f(\Gv{}_1) \geq \alpha_{GW} \cdot \textsc{OPT}_{SDP}$ in expectation, where $\alpha_{GW} \geq 0.878567$.

\begin{mybox2}
\begin{lemma}\label{lemma:sdp-quality}
$\mathbb{E}[f(\Gv{}_1)] \geq \alpha_{GW} \cdot \textsc{OPT}_{SDP}$ where $\alpha_{GW} \geq 0.878567$.
\end{lemma}
\end{mybox2}

\begin{proof}
By Lemma~(\ref{lemma:GW}), for any two vertices $i$ and $j$, let $H_{ij}$ be the event where $i$ and $j$ are in the same subset.
we have 
\begin{equation}\label{eq:Hij}
    \Pr[H_{ij}] = 1 - \frac{\arccos{} (\vec{y}_i \cdot \vec{y}_j)}{\pi}
\end{equation}

\noindent
Let $\mc{P}$ be the \assignment{} where we assign type-\texttt{1} (type-\texttt{2}) to vertices in $\Gv{}_1$ ($\Gv{}_2$). since a vertex $i$ is not integrated if and only if $i$ and all its neighbors are in the same set, we have
\begin{align*}
    \Pr[\text{vertex } i \text{ is integrated}] &= 1 - \Pr[\text{vertex } i \text{ is not integrated}]\\
    &= 1 - \Pr[\bigcap_{j \in \mc{N}(i)} H_{ij}]
\end{align*}

Further,
\begin{align*}
    \Pr[\bigcap_{j \in \mc{N}(i)} H_{ij}] &\leq \min_{j \in \mc{N}(i)} \{ \; \Pr[H_{ij}] \; \}\\
    &= \min_{j \in \mc{N}(i)} \{ 1 - \frac{\arccos{} (\vec{y}_i \cdot \vec{y}_j)}{\pi} \} \;\;\;\;\;\;\;\; (\text{By Eq.}~(\ref{eq:Hij}))\\
    &= 1 - \max_{j \in \mc{N}(i)} \{ \; \frac{\arccos{} (\vec{y}_i \cdot \vec{y}_j)}{\pi} \; \}
\end{align*}

\noindent
and thus
\begin{equation}
    \Pr[i \text{ is integrated}] \geq \max_{j \in \mc{N}(i)} \{ \; \frac{\arccos{} (\vec{y}_i \cdot \vec{y}_j)}{\pi} \; \}
\end{equation}

As show in~\cite{goemans1995improved}, $\arccos{}(z) / \pi \geq \alpha_{GW} \cdot (1 - z) / 2$ for real $z \in [-1, 1]$. It follows that
\begin{equation}
    \max_{j \in \mc{N}(i)} \{ \; \frac{\arccos{} (\vec{y}_i \cdot \vec{y}_j)}{\pi} \; \} \geq \alpha_{GW} \cdot \max_{j \in \mc{N}(i)} \{\frac{1 - \vec{y}_i \cdot \vec{y}_j}{2}\}
\end{equation}
since $\vec{y}_i \cdot \vec{y}_j \in [-1, 1]$. Recall that $f(\Gv{}_1)$ is the number of integrated vertices under the assignment where type-\texttt{1} (type-\texttt{2})  are assigned to $\Gv{}_1$ ($\Gv{}_2$). We have
\begin{align}
    \mb{E}[f(\Gv{}_1)] &\geq \sum_{i \in \Gv{}} \max_{j \in \mc{N}(i)} \{ \; \frac{\arccos{} (\vec{y}_i \cdot \vec{y}_j)}{\pi} \; \}\\
    &\geq \alpha_{GW} \cdot \sum_{i \in \Gv{}} \max_{j \in \mc{N}(i)} \{\frac{1 - \vec{y}_i \cdot \vec{y}_j}{2}\}\\
    &\geq \alpha_{GW} \cdot \textsc{OPT}_{SDP}
\end{align}
This concludes the proof.
\end{proof}

\subsubsection*{Second step: Fix the size}
In the previous step, we have shown that given a partition $\{\Gv{}_1, \Gv{}_2\}$ resulted from hyperplane rounding, if all vertices in $\Gv{}_1$ are of type-\texttt{1}, and all vertices in $\Gv{}_2$ are of type-\texttt{2}, then the expected number of integrated vertices is at least $\alpha_{GW}$ of the optimal. Nevertheless, there is one problem: {\em the partition is not necessarily an $(\alpha n, 1- \alpha) n$-partition.} In this section, we present an algorithm to move vertices from one subset to another such that $(i)$ the resulting new partition is an $(\alpha n, (1- \alpha) n)$-partition, and $(ii)$ the objective does not decrease ``too much'' after the moving process.

\begin{algorithm}
\caption{\texttt{Fix-Size}}\label{alg:algo2-move}
\SetKwInOut{Input}{Input}
\SetKwInOut{Output}{Output}

\Input{Subsets $\Gv{}_1$ and $\Gv{}_2$}
\Output{Subsets $\Gv{}_1^{(T)}$ and $\Gv{}_2^{(T)}$, where $T = |\Gv{}_1| - \alpha n$}

$\Gv{}_1^{(0)} \gets \Gv{}_1$, $\Gv{}_2^{(0)} \gets \Gv{}_2$

\For{$t$ from $1$ to $T$}
{
  $i \gets \argmax_{j \in \Gv{}_1^{(t-1)}} \{f(\Gv{}_1^{(t-1)} \setminus \{j\}) - f(\Gv{}_1^{(t-1)})\}$   
  
  $\Gv{}_1^{(t)} \gets \Gv{}_1^{(t-1)} \setminus \{i\}$
  
  $\Gv{}_2^{(t)} \gets \Gv{}_2^{(t-1)} \cup \{i\}$
}
\Return{$\{\Gv{}_1^{(T)}, \Gv{}_2^{(T)}\}$}
\end{algorithm}

\paragraph{Algorithm for the second step.} Without losing generality, suppose $|\Gv{}_1| \geq \alpha n$. Overall, our algorithm consists of $T = |\Gv{}_1| - \alpha n$ iterations, and in each each iteration, we move a vertex $i \in \Gv{}_1$ to $\Gv{}_2$. Specifically, let $\Gv{}_1^{(t)}$ be the subset at the $t$th iteration, with $\Gv{}_1^{(0)} = \Gv{}_1$. To obtain $\Gv{}_1^{(t+1)}$, we choose $i \in \Gv{}_1^{(t)}$ to be a vertex that maximizes $f(\Gv{}_1^{(t)} \setminus \{i\}) - f(\Gv{}_1^{(t)})$, and the move $i$ to the other subset. A pseudocode is given in Algorithm~(\ref{alg:algo2-move}). 

\noindent
Let $\Gv{}_1^{(T)}$, $T = |\Gv{}_1| - \alpha n$, be the subset returned by Algorithm~(\ref{alg:algo2-move}).

\begin{mybox2}
\begin{lemma}\label{lem:average-increase}
We have 
$$
\frac{f(\Gv{}_1^{(T)})}{|\Gv{}_1^{(T)}|} \geq \frac{f(\Gv{}_1)}{|\Gv{}_1|}
$$
\end{lemma}
\end{mybox2}

\begin{proof}
For each vertex $j \in \Gv{}_1^{(t)}$, $0 \leq t \leq T-1$, let $\eta_j^{(t)}$ be the number of its neighbors in $\Gv{}_2^{(t)}$ that are not adjacent to any other vertices in $\Gv{}_1^{(t)}$. Further, let $\nu^{(t)}$ be the number of vertices in $\Gv{}_2^{(t)}$ that have more than one neighbor in $\Gv{}_1^{(t)}$, and let $\delta^{(t)}$ be the number of vertices in $\Gv{}_1^{(t)}$ that has at least one neighbor in $\Gv{}_2^{(t)}$. We then have that $f(\Gv{}_1^{(t)}) = \delta^{(t)} + \nu^{(t)} + \sum_{j \in \Gv{}_1^{(t)}} \eta_j^{(t)}$, thus 
$$
    \frac{f(\Gv{}_1^{(t)})}{|\Gv{}_1^{(t)}|} = \frac{\delta^{(t)} + \nu^{(t)} + \sum_{j \in \Gv{}_1^{(t)}} \eta_j^{(t)}}{|\Gv{}_1^{(t)}|}
$$

\noindent
We now argue that
\begin{equation}\label{eq:mean-increase}
    \frac{f(\Gv{}_1^{(t)})}{|\Gv{}_1^{(t)}|} \leq \frac{f(\Gv{}_1^{(t+1)})}{|\Gv{}_1^{(t+1)}|}
\end{equation}
Note that if there exists at least one vertex in $\Gv{}_1^{(t)}$ that has no neighbors in $\Gv{}_2^{(t)}$ (i.e., $\delta^{(t)} < |\Gv{}_1^{(t)}|$), then such a vertex will be chosen. One can easily verify that the resulting new objective $f(\Gv{}_1^{(t+1)})$ is greater than $f(\Gv{}_1^{(t)})$ and the above inequality~(\ref{eq:mean-increase}) clearly holds. 

\par Now suppose that all vertices in $\Gv{}_1^{(t)}$ have neighbors on the other side. Note that after moving any vertex $j$ from $\Gv{}_1^{(t)}$ to $\Gv{}_2^{(t)}$, the decrease of the objective is at most $\eta_{j} + 1$, where the additional plus one comes from the possibility of $j$ itself becoming non-integrated. By the greedy nature of the algorithm, we have that $f(\Gv{}_1^{(t)}) - f(\Gv{}_1^{(t+1)}) \leq \eta_{min} + 1$, where $\eta_{min} = \min_j \{\eta_j\}$. It follows that 
\begin{align}\label{eq:fixing}
    \frac{f(\Gv{}_1^{(t+1)})}{|\Gv{}_1^{(t+1)}|} &\geq \frac{f(\Gv{}_1^{(t)}) - \eta_{min} - 1 }{|\Gv{}_1^{(t)}| - 1} \nonumber\\
    &= \frac{\delta^{(t)} - 1}{|\Gv{}_1^{(t)}| - 1} + \frac{\nu^{(t)}}{|\Gv{}_1^{(t)}| - 1} + \frac{\left(\sum_{j \in \Gv{}_1^{(t)}} \eta_j^{(t)}\right) - \eta_{min}}{|\Gv{}_1^{(t)}| - 1} \nonumber\\
    &\geq \frac{\delta^{(t)}}{|\Gv{}_1^{(t)}|} + \frac{\nu^{(t)}}{|\Gv{}_1^{(t)}|} + \frac{\left(\sum_{j \in \Gv{}_1^{(t)}} \eta_j^{(t)}\right)}{|\Gv{}_1^{(t)}|}\\
    &= \frac{f(\Gv{}_1^{(t)})}{|\Gv{}_1^{(t)}|} \nonumber
\end{align}
Lastly, by recursion, we have that 
\begin{equation}
    \frac{f(\Gv{}_1^{(T)})}{|\Gv{}_1^{(T)}|} \geq \frac{f(\Gv{}_1^{(0)})}{|\Gv{}_1^{(0)}|} =  \frac{f(\Gv{}_1)}{|\Gv{}_1|}
\end{equation}
This concludes the proof.
\end{proof}

\paragraph{The final algorithm.} We have defined the two steps (i.e., $(i)$ {\em round the SDP} and $(ii)$ {\em fix the sizes of the two subsets}) that we need to take to obtain a feasible solution of the problem. Let $\epsilon \geq 0$ be a small constant, and let $L = \ceil{\log_{a}({\frac{1}{\epsilon}})}$ where $a = [(1 + \beta) - (1 - \epsilon) 2 \alpha_{GW}] / (1 + \beta - 2 \alpha_{GW})$, $\beta = 1/(4(\alpha - \alpha^2))$. Note that $L$ is a constant w.r.t. $n$. The final algorithm consists of $L$ iterations, where each iteration performs the two steps defined above. This gives us $L$ feasible solutions. The algorithm then outputs a solution with the highest objective.

\subsubsection*{Analysis of the final algorithm}

\begin{mybox2}
\begin{theorem}
    The final algorithm gives a factor 
    $$\frac{\alpha \left( (1 - \epsilon) \cdot 2\alpha_{GW} - \frac{\gamma-\gamma^2}{\alpha-\alpha^2} \right)}{\gamma} \cdot (1 - \epsilon)$$ 
    approximation w.h.p. where $\alpha_{GW} \geq 0.878567$, $\epsilon \geq 0$ is an arbitrarily small positive constant, $\alpha = k/n$ is the fraction of minority agents in the group, and $\gamma = \sqrt{\alpha (1 - \alpha) (1 - \epsilon) \cdot 2 \alpha_{GW}}$.
\end{theorem}
\end{mybox2}

\begin{proof}
The analysis of the final algorithm follows a same route as the one in~\cite{frieze1997improved}. For any iteration $1 \leq \ell \leq L$ of the algorithm, let $\Gv{}_1$ and $\hat{\Gv{}}_1$ be the subsets returned after the first step and the second step, respectively. In Lemma~(\ref{lem:average-increase}), we have shown that 
\begin{equation}
    \frac{f(\hat{\Gv{}_1})}{|\hat{\Gv{}_1}|} \geq \frac{f(\Gv{}_1)}{|\Gv{}_1|} \cdot
\end{equation}

Let $X = f(\Gv{}_1)$ be a random variable denoting the objective of the solution after the rounding, before performing the second step. Let $Y = |\Gv{}_1| \cdot |\Gv{}_2|$ be another random variable, representing the product of the sizes of the two partitions. Lemma~(\ref{lemma:GW}) have shown that $\mathbb{E}[X] \geq \alpha_{GW} \cdot \textsc{OPT}_{SDP}$. For the expected value of $Y$, we have
\begin{align}
    \mathbb{E}[Y] &= \sum_{i < j} \Pr [i \text{ and } j \text{ are in different subset}] \nonumber\\
    &\geq \alpha_{GW} \cdot \sum_{i < j} \frac{1 - \vec{y}_i \cdot  \vec{y}_j}{2}
\end{align}

where the second inequality follows from Lemma~(\ref{lemma:GW}). By the SDP constraint $\sum_{i < j} \vec{y}_i \cdot \vec{y}_j \leq [(1 - 2 \alpha)^2 \cdot n^2 - n] / 2$, we can further show that
\begin{align}
    \sum_{i < j}1 - \vec{y}_i \cdot \vec{y}_j &= \frac{n(n-1)}{2} -  \sum_{i < j} \vec{y}_i \cdot \vec{y}_j \nonumber\\
    &\geq \frac{n(n-1)}{2} - \frac{(1 - 2 \alpha)^2 \cdot n^2 - n}{2}\\
    &= \frac{\left( (2 - 2 \alpha) \cdot 2 \alpha \right) \cdot n^2}{2} \nonumber
\end{align}
It follows that 
\begin{align}
    \mathbb{E}[Y] \geq \alpha_{GW} \cdot \frac{\left( (2 - 2 \alpha) \cdot 2 \alpha \right) \cdot n^2}{4} = \alpha_{GW} \cdot (\alpha - \alpha^2) \cdot n^2
\end{align}
Let $N = (\alpha - \alpha^2) \cdot n^2$. Let random variable $Z = X / \textsc{OPT}_{SDP} + Y/N$, then $\mathbb{E}[Z] \geq 2 \cdot \alpha_{GW}$. Further, since $Y \leq n^2 / 4$, one can verify that $Y/N \leq 1/(4(\alpha - \alpha^2))$. Overall, we have $Z \leq 1 + \beta$ where $\beta = 1/(4(\alpha - \alpha^2))$.

\noindent
With simple Markov inequality, we have
\begin{equation}
    \Pr[Z \leq (1 - \epsilon) \cdot 2 \alpha_{GW}] \leq \frac{(1 + \beta) - 2 \alpha_{GW}}{(1 + \beta) - (1 - \epsilon) \cdot 2 \alpha_{GW}}
\end{equation}
Note that there is a random variable $Z$ for each of the iteration. Define $Z'$ to be the largest $Z$ over all the $L$ iterations where $Z' = X'/\textsc{OPT}_{SDP} + Y'/N$. It follows that
\begin{equation}
    \Pr[Z' \leq (1 - \epsilon) \cdot 2 \alpha_{GW}] \leq \left(\frac{(1 + \beta) - 2 \alpha_{GW}}{(2 + \beta) - (1 - \epsilon) \cdot 2 \alpha_{GW}}\right)^{L} \leq \epsilon
\end{equation}
For our choice of $L = \ceil{\log_{a}(1/\epsilon)}$ where $a = [(1 + \beta) - (1 - \epsilon) \cdot 2 \alpha_{GW}] / (1 + \beta - 2 \alpha_{GW}) $. We consider the case where
$$
Z' \geq (1 - \epsilon) \cdot 2 \alpha_{GW}
$$
which happens with probability at least $1 - \epsilon$. Let $\rho = X'/\textsc{OPT}_{SDP}$ be the ratio of $X'$ to the optimal of SDP. Then by the definition $Z' = X'/\textsc{OPT}_{SDP} + Y'/N$, one can verify that 
\begin{equation}\label{eq:Y'}
   Y' \geq \left( (1 - \epsilon) \cdot 2  \alpha_{GW} - \rho \right) N 
\end{equation}
Let $\mu = |\Gv{}_1'| \; / \; n$ where $\Gv{}_1'$ is the subset for type-\texttt{1} vertices obtained after the first step (i.e., rounding) in the iteration for $Z'$. Then $Y' = \mu(1-\mu) n^2$. Using equation~(\ref{eq:Y'}) and $N = (\alpha - \alpha^2) n^2$, we can verify that
\begin{equation}\label{eq:rho}
    \rho \geq (1 - \epsilon) \cdot 2 \alpha_{GW} - \frac{\mu-\mu^2}{\alpha-\alpha^2}
\end{equation}
Now let $\hat{\Gv{}}_1'$ be the subset of type-\texttt{1} vertices after fixing the size of $\Gv{}_1'$ (i.e., after the second step). Based on Lemma~(\ref{lem:average-increase}) and equation~(\ref{eq:rho}), we have
\begin{align}
    f(\hat{\Gv{}}_1') &\geq \frac{f(\Gv{}_1')}{|\Gv{}_1'|} \cdot |\hat{\Gv{}}_1'|\\
    &= \frac{f(\Gv{}_1') \cdot \alpha}{\mu}\\
    &= \frac{\alpha \cdot \rho}{\mu} \cdot \textsc{OPT}_{SDP}\\
    &\geq \frac{\alpha \left( (1 - \epsilon) \cdot 2 \alpha_{GW} - \frac{\mu-\mu^2}{\alpha-\alpha^2} \right)}{\mu} \cdot \textsc{OPT}_{SDP}
\end{align}
One can verify that for $\mu > 0$, $\frac{\alpha \left( (1 - \epsilon) \cdot 2 \alpha_{GW} - \frac{\mu-\mu^2}{\alpha-\alpha^2} \right)}{\mu}$ is minimum at $\mu = \sqrt{\alpha (1 - \alpha) (1 - \epsilon) \cdot 2 \alpha_{GW}}$. 

Let $\gamma = \sqrt{\alpha (1 - \alpha) (1 - \epsilon) \cdot 2 \alpha_{GW}}$, we then have
\begin{equation}
    f(\hat{\Gv{}}_1') \geq \frac{\alpha \left( (1 - \epsilon) \cdot 2 \alpha_{GW} - \frac{\gamma-\gamma^2}{\alpha-\alpha^2} \right)}{\gamma} \cdot \textsc{OPT}_{SDP}
\end{equation}
Lastly, since $\Pr[Z' > (1-\epsilon) \cdot 2 \alpha_{GW}] \geq 1 - \epsilon$,
\begin{equation}
    \mathbb{E}[f(\hat{\Gv{}}_1')] \geq \frac{\alpha \left( (1 - \epsilon) \cdot 2 \alpha_{GW} - \frac{\gamma-\gamma^2}{\alpha-\alpha^2} \right)}{\gamma} \cdot (1 - \epsilon) \cdot \textsc{OPT}_{SDP}
\end{equation}
\end{proof}

For small enough $\epsilon$, say $\epsilon = 10^{-3}$, the approximation ratio is greater than $1/2$ for $\alpha$ in range $[0.4, 0.5]$. For example, $\alpha = 0.45$ gives a ratio of $0.5781$, and $\alpha = 0.5$ gives a ratio of $0.6492$.

\section{Additional Material for Section 6}
In this section, we first show that \maxinte{}-\IoAgent{} can be solved in polynomial time on treewidth bounded graphs. Based on this result, we further present a {\em polynomial time approximation scheme} (PTAS) for the problem on planar graphs.

\subsection{A dynamic programming algorithm for treewidth bounded graphs}
The concept {\em treewidth} of a graph is first introduced in the seminal work by Robertson and Seymour~\cite{robertson1986graph}. Many intractable problems have since enjoyed polynomial time algorithms when underlying graphs have bounded treewidth. In this section, we present a dynamic programming algorithm that solves \maxinte{}-\IoAgent{} in polynomial time (w.r.t. $n$) for the class of graphs that are treewidth bounded. 

\paragraph{Dynamic programming setup.} Given an instance of \maxinte{}-\IoAgent{} with graph $\G = (\Gv{}, \Ge{})$ and the number $k$ of minorities, let $\mc{T} = (\mc{I}, \mc{F})$ be a tree decomposition of $\G{}$ with a bounded treewidth $\sigma$. For each $\mc{X}_i \in \mc{I}$, consider the set of bags in the subtree rooted at $\mc{X}_i$ in $\mc{T}$, and let $\mc{Y}_i$ be the set of all vertices in these bags. Let $\G{}[\mc{Y}_i]$ denote the graph $\G{}$ induced on $\mc{Y}_i$. For each bag $\mc{X}_i$, we define an array $H_i$ to keep track of the optimal objectives in $\G{}[\mc{Y}_i]$.

\begin{figure}[!h]
  \centering
    \includegraphics[width=0.7\textwidth]{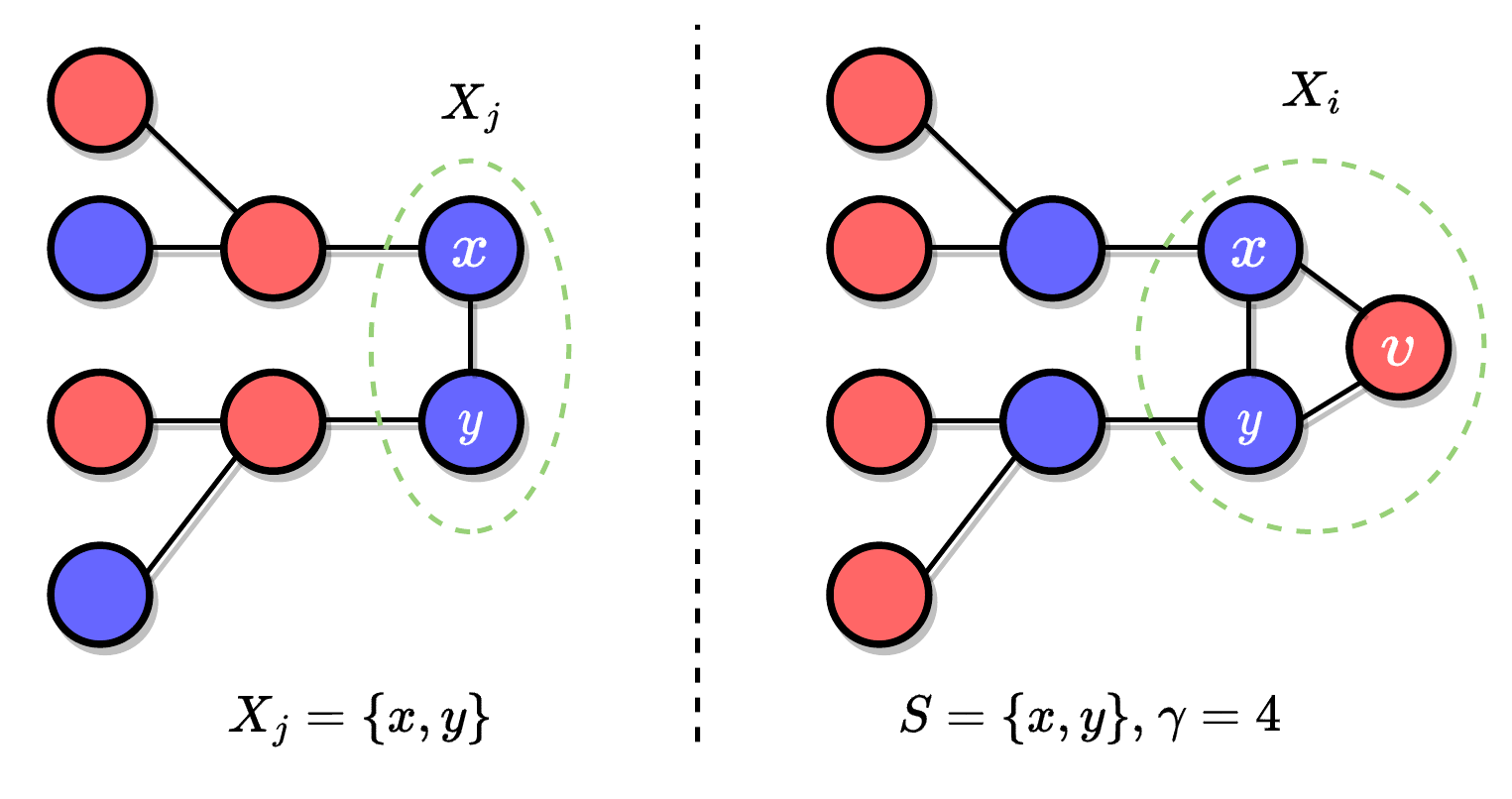}
    \caption{An example where the naive dp approach fails. Given an example graph $\G{}$, the subgraph on the left is $\G{}$ induced on $\mc{Y}_j$, where $\mc{X}_j = \{x, y\}$. The subgraph on the right is $\G{}$ induced on $\mc{Y}_i$, where $\mc{X}_j = \{x, y, v\}$. The set $S = \{x, y\}$, and $\gamma = 4$. Optimal assignments that yields $H_j(S, \gamma)$ and $H_i(S, \gamma)$ are given where blue vertices are of type-\texttt{1}, and red vertices are of type-\texttt{2}. In particular, $H_j(S, \gamma) = 6$ and $H_i(S, \gamma) = 9$. Note that the naive dp approach would set $H_i(S, \gamma)$ to be $6 + 1 = 7$ which is not optimal.} 
    \label{fig:naive_dont_work}
\end{figure}

\paragraph{A naive definition that fails.} One immediate way is to define $H_i(S, \gamma)$ to be the optimal objective in $\G{}[\mc{Y}_i]$ such that $(i)$ $S \subseteq \mc{X}_i$ are of type-\texttt{1}, $\mc{X}_i \setminus S$ are of type-\texttt{2}; $(ii)$ there are a total of $\gamma$ type-\texttt{1} vertices and $|\mc{Y}_i| - r$ type-\texttt{2} vertices in $\G{}[\mc{Y}_i]$. As a result, for each $\mc{X}_i$, its corresponding $H_i$ has $O(2^{\sigma} \cdot n)$ entries, which is polynomial w.r.t. $n$ since $\sigma$ is bounded. Despite the simplicity of this definition, however, it is unclear how to correctly update these arrays. For example, suppose $\mc{X}_i$ is of the type \textsf{introduce}, let $\mc{X}_j$ be the child of $\mc{X}_i$. Let $v \notin S$ be the vertices that is introduced to $\mc{X}_i$. One might try to update $H_i(S, \gamma)$ by doing $H_i(S, \gamma) = H_j(S, \gamma) + w(v, \mc{X}_i)$, where $w(v, \mc{X}_i)$ is the number of newly integrated vertices in $\mc{X}_i$ after $v$ being introduced to the set. This formulation looks correct at the first glance since $v$ is not adjacent to any vertices in $\mc{Y}_i$ other than those in $\mc{X}_i$. Thus, it seems that the impact this extra vertex $v$ can cause is only restricted within $\mc{X}_i$. However, we remark this far from true, and that the above computation is not optimal. In particular, consider the example given in Fig~(\ref{fig:naive_dont_work}).

\paragraph{An alternative definition.} We introduce another dimension to the above definition of $H_i$. In particular, let $H_i(S, S', \gamma)$ be the optimal objective in $\G{}[\mc{Y}_i]$ such that 
\begin{itemize}
    \item[$(i)$] Vertices in the subset $S \subseteq \mc{X}_i$ are of type-\texttt{1}, and vertices in $\mc{X}_i \setminus S$ are of type-\texttt{2}.
    \item[$(ii)$] Vertices in $S' \subseteq \mc{X}_i$ are to be {\em treated integrated}. 
    \item[$(iii)$] There is a total of $\gamma$ type-\texttt{1} vertices and $|\mc{Y}_i| - \gamma$ type-\texttt{2} vertices in $\G{}[\mc{Y}_i]$.
\end{itemize}

\noindent
The resulting $H_i$ has $O(4^{\sigma} \cdot n)$ entries. The algorithm then proceeds in a bottom-up fashion from the leaves to the root in $\mc{T}$. We now discuss how the array $H_i$ is updated for each bag $\mc{X}_i$. 

\subsubsection*{Update Scheme}

\begin{itemize}
    \item[$(i)$] \textbf{\textsf{Leaf:}} For all $\gamma = 0, ..., \min\{|\mc{Y}_i|, k\}$ (recall that $k$ is the total number of type-\texttt{1} agents) and for all $S \subseteq \mc{X}_i$ s.t. $|S| = \gamma$, let $Z_i(S)$ be the set of integrated vertices in $\G{}[\mc{X}_i]$ under the assignment $(S, 
    \mc{X}_i \setminus S)$. For all $S' \subseteq X_i$, we have
    \begin{equation}
        H_i(S, S', \gamma) = |S' \cup Z_i(S)|
    \end{equation}
    
    \item[$(ii)$] \textbf{\textsf{Introduce:}} Let $\mc{X}_j$ be the child of $\mc{X}_i$, and let $v$ be the vertex introduced to $\mc{X}_i$ (i.e., $v \in \mc{X}_i$ and $v \notin \mc{X}_j$). For all $\gamma = 0, ..., \min\{|\mc{Y}_i|, k\}$, and for all $S \subseteq \mc{X}_i$ s.t. $|S| \leq \gamma$, let $Z_i(S)$ be the set of integrated vertices in $\G{}[\mc{X}_i]$ under the assignment $(S, \mc{X}_i \setminus S)$. For all $S' \subseteq \mc{X}_i$:
    \begin{itemize}
        \item[-] If $v \in S$,
        \begin{equation}
            H_i(S, S', \gamma) = H_j(S \setminus \{v\}, S' \cup Z_i(S) \setminus \{v\}, \gamma - 1) + \mathbbm{1}(v)
        \end{equation}
        \item[-] If $v \notin S$,
            \begin{equation}
            H_i(S, S', \gamma) = H_j(S, S' \cup Z_i(S) \setminus \{v\}, \gamma) + \mathbbm{1}(v)
        \end{equation}
    \end{itemize}
    
    where $\mathbbm{1}(v)$ is an indicated variable that equals to $1$ if and only if $v$ is integrated in $\G{}[\mc{X}_i]$ under the assignment $(S, \mc{X}_i \setminus S)$.
    
    \item[$(iii)$] \textbf{\textsf{Forget:}} Let $\mc{X}_j$ be the child of $\mc{X}_i$, and let $v$ be the vertex forgot by $\mc{X}_i$ (i.e., $v \notin \mc{X}_i$ and $v \in \mc{X}_j$). For all $\gamma = 0, ..., \min\{|\mc{Y}_i|, k\}$, and for all $S \subseteq \mc{X}_i$ s.t. $|S| \leq \gamma$, let $Z_i(S)$ be the set of integrated vertices in $\G{}[\mc{X}_i]$ under the assignment $(S, \mc{X}_i \setminus S)$. For all $S' \subseteq \mc{X}_i$:
    \begin{equation}
        H_i(S, S', \gamma) = \max\{H_j(S, Z_i(S) \cup S', \gamma), H_j(S \cup \{v\}, Z_i(S) \cup S', \gamma)\}
    \end{equation}
    
    \item[$(iv)$] \textbf{\textsf{Join:}} Let $\mc{X}_{j1}$ and $\mc{X}_{j2}$ be the two children of $\mc{X}_i$. Note that $\mc{X}_{j1} = \mc{X}_{j2} = \mc{X}_{i}$. For all $\gamma = 0, ..., \min\{|\mc{Y}_i|, k\}$, and for all $S \subseteq \mc{X}_i$ s.t. $|S| \leq \gamma$, let $Z_i(S)$ be the set of truly integrated vertices in $\G{}[\mc{X}_i]$ under the assignment $(S, \mc{X}_i \setminus S)$.
    
    \par For all $S' \subseteq \mc{X}_i$, let $Q_i(S, S') = \mc{X}_i \setminus \left( S' \cup Z_i(S) \right)$ be the set of vertices that are \textbf{not} truly integrated in $\G{}[\mc{X}_i]$ under the assignment $(S, \mc{X}_i \setminus S)$, and also should \textbf{not} be treated as integrated (i.e., vertices in $Q_i(S, S')$ are not in $S'$). We consider all subsets $Q_{j1}(S, S') \subseteq Q_i(S, S')$ and $Q_{j2}(S, S') \subseteq Q_i(S, S')$, let $\Bar{Q}_{j1}(S, S') = Q_i(S, S') \setminus Q_{j1}(S, S')$ and $\Bar{Q}_{j2}(S, S') = Q_i(S, S') \setminus Q_{j2}(S, S')$.

    \par Consider the solutions $H_{j1}(S, S' \cup Z_i(S) \cup \Bar{Q}_{j1}(S, S'), \gamma_1)$ and $H_{j2}(S, S' \cup Z_i(S) \cup \Bar{Q}_{j2}(S, S'), \gamma_2)$. Let $\pla{}[Q_{j1}(S, S'), \gamma_1]$ and $\pla{}[Q_{j2}(S, S'), \gamma_2]$ be two corresponding assignments, restricted to $\mc{Y}_{j1}$ and $\mc{Y}_{j2}$, that yield the objective $H_{j1}(S, S' \cup Z_i(S) \cup \Bar{Q}_{j1}(S, S'), \gamma_1)$ and $H_{j2}(S, S' \cup Z_i(S) \cup \Bar{Q}_{j2}(S, S'), \gamma_2)$, respectively. Such assignments can be easily obtained during the bottom-up process. 
    
    Lastly, let $W(\pla{}[Q_{j1}(S, S'), \gamma_1])$ and $W(\pla{}[Q_{j2}(S, S'), \gamma_2])$ be the set of \textbf{truly} integrated vertices in $\mc{Y}_{j1} \setminus \left( S' \cup Z_i(S) \right)$ and $\mc{Y}_{j2} \setminus \left( S' \cup Z_i(S) \right)$ under $\pla{}[Q_{j1}(S, S'), \gamma_1]$ and $\pla{}[Q_{j2}(S, S'), \gamma_2]$, respectively. $H_i(S, S', \gamma)$ is \textbf{computed as follows}:
    \begin{equation}\label{eq:join}
        H_i(S, S', \gamma) = \max_D\{ |W(\pla{}[Q_{j1}(S, S'), \gamma_1]) \cup W(\pla{}[Q_{j2}(S, S'), \gamma_2])| \} + |S' \cup Z_i(S)|
    \end{equation}

\noindent
    where 
    \begin{align*}
    D = \{\gamma_1, \gamma_2, Q_{j1}(S, S'), Q_{j2}(S, S') &: \gamma_1 + \gamma_2 - |S| = \gamma, \gamma_1 \geq |S|, \gamma_2 \geq |S|, \\ 
    &Q_{j1}(S, S') \subseteq Q_i(S, S'), \; Q_{j2}(S, S') \subseteq Q_i(S, S')\}
    \end{align*}
    is the set of variables. 
\end{itemize}

\begin{mybox2}
\begin{theorem}
The problem \maxinte{}-\IoAgent{} can be solved optimally in polynomial time on tree-width bounded graphs.
\end{theorem}
\end{mybox2}

\begin{proof}
We first analyze the correctness of the update scheme. The optimality of the first three update rules (i.e., \textsf{leaf}, \textsf{introduce}, \textsf{forget}) easily follows from induction. We further discuss the case where $\mc{X}_i$ is of type \textsf{join}. In particular, we argue that the optimal objective $H_i(S, S', \gamma)$ has the value shown in Equation~(\ref{eq:join}).

\par Consider an optimal assignment $\plas{}_i$ on $\G{}[\mc{Y}_i]$ that achieves the optimal objective $H_i(S, S', \gamma)$. Let $\gamma_1^*$ and $\gamma_2^*$ be the number of type-1 vertices in $G[\mc{Y}_{j1}]$ and in $G[\mc{Y}_{j2}]$, respectively, under $\plas{}_i$. Since $\mc{Y}_{j1}$ and $\mc{Y}_{j2}$ only share $\mc{X}_i$ as a common set, we have $\gamma_1^* + \gamma_2^* - |S| = \gamma$. We may consider $\plas{}_i$ as a union of two assignments, $\plas{}_{j1}$ and $\plas{}_{j2}$, where $\plas{}_{j1}$ and $\plas{}_{j2}$ are $\plas{}_i$ restricted to $G[\mc{Y}_{j1}]$ and in $G[\mc{Y}_{j2}]$, respectively.

\par Recall that $Q_i(S, S') = \mc{X}_i \setminus \left( S' \cup Z_i(S) \right)$ is the set of vertices that are \textbf{not} truly integrated in $\G{}[\mc{X}_i]$ under the assignment $(S, \mc{X}_i \setminus S)$, and also should \textbf{not} be treated as integrated (i.e., vertices in $Q_i(S, S')$ are not in $S'$). Note that it is possible that some vertices in $Q_i(S, S')$ are integrated under $\plas{}_{j1}$ and $\plas{}_{j1}$. In particular, let $Q_{j1}^*(S, S') \subseteq Q_i(S, S')$ and $Q_{j2}^*(S, S') \subseteq Q_i(S, S')$ be the set of vertices in $Q_i(S, S')$ that are good under $\plas{}_{j1}$ and $\plas{}_{j2}$, respectively. 

Observe that $\gamma_1^*, \gamma_2^*, Q_{j1}^*(S, S'), Q_{j2}^*(S, S') \in D$. Let $\pla{}[Q_{j1}^*(S, S'), \gamma_1^*]$ and $\pla{}[Q_{j2}^*(S, S'), \gamma_2^*]$ be two corresponding assignments returned by the proposed update scheme, restricted to $\mc{Y}_{j1}$ and $\mc{Y}_{j2}$, that yield the objective $H_{j1}(S, S' \cup Z_i(S) \cup \Bar{Q}^*_{j1}(S, S'), \gamma_1^*)$ and $H_{j2}(S, S' \cup Z_i(S) \cup \Bar{Q}^*_{j2}(S, S'), \gamma_2^*)$, respectively. In particular, we have
\begin{align}
    \IoAgent{}(\pla{}[Q_{j1}^*(S, S'), \gamma_1^*]) &= H_{j1}(S, S' \cup Z_i(S) \cup \Bar{Q}^*_{j1}(S, S'), \gamma_1^*)\\ &= |S' \cup Z_i(S) \cup \Bar{Q}^*_{j1}(S, S')| + |W(\pla{}[Q_{j1}(S, S'), \gamma_1]) \setminus \mc{X}_{j1}|\\ &+ | W(\pla{}[Q_{j1}(S, S'), \gamma_1]) \cap Q_{j1}^*(S, S')|
\end{align}
Consider the particular instance $(S, S' \cup Z_i(S) \cup \Bar{Q}^*_{j1}(S, S'), \gamma_1^*)$ for $\G{}[\mc{Y}_{j1}]$. 
Since $\pla{}[Q_{j1}^*(S, S'), \gamma_1^*]$ is an optimal solution and $\plas{}_{j1}$ is a feasible solution of this instance, it follows that 
\begin{align}
    |W(\plas{}_{j1}) \setminus \mc{X}_{j1}| + |W(\plas{}_{j1}) \cap Q_{j1}^*(S, S')| &= |W(\plas{}_{j1})|\\
    & \leq |W(\pla{}[Q_{j1}(S, S'), \gamma_1]) \setminus \mc{X}_{j1}|\\ &+ | W(\pla{}[Q_{j1}(S, S'), \gamma_1]) \cap Q_{j1}^*(S, S')|
\end{align}
Similarly, we also have
\begin{equation}
    |W(\plas{}_{j2})|
     \leq |W(\pla{}[Q_{j2}(S, S'), \gamma_2]) \setminus \mc{X}_{j2}| + | W(\pla{}[Q_{j2}(S, S'), \gamma_2]) \cap Q_{j2}^*(S, S')|
\end{equation}

\noindent
The objective of $\plas{}_i$ for the instance $(S, S', \gamma)$ on $\mc{X}_i$ is of the form:
\begin{align}
    \IoAgent{}(\plas{}_i) &= H_i(S, S', \gamma)\\
    &= |S' \cup Z_i(S)| + |W(\plas{}_{j1}) \cup W(\plas{}_{j2})|\\ 
    &= |S' \cup Z_i(S)| + |W(\plas{}_{j1})| + |W(\plas{}_{j2})| - |Q_{j1}^*(S, S') \cap Q_{j2}^*(S, S')|
\end{align}

\noindent
Consider the placement $\pla{}_i$ which is the union of $\pla{}[Q_{j1}^*(S, S'), \gamma_1^*]$ and $\pla{}[Q_{j2}^*(S, S'), \gamma_2^*]$. Note that $\pla{}_i$ is a feasible solution to the problem instance $(S, S', \gamma)$ on $\mc{X}_i$ since $\gamma_1^* + \gamma_2^* - |S| = \gamma$, and $\mc{Y}_{j1}$ only overlaps with $\mc{Y}_{j1}$ on $\mc{X}_i$. The objective of $\pla{}_i$ for the instance $(S, S', \gamma)$ on $\mc{X}_i$ satisfies the following inequality:
\begin{align}
    \IoAgent{}(\pla{}_i) &= |S' \cup Z_i(S)| + |W(\pla{}[Q_{j1}(S, S'), \gamma_1]) \cup W(\pla{}[Q_{j2}(S, S'), \gamma_2])|\\ 
    &\geq |S' \cup Z_i(S)|\\ &+|W(\pla{}[Q_{j1}(S, S'), \gamma_1]) \setminus \mc{X}_{j1}| + |W(\pla{}[Q_{j1}(S, S'), \gamma_1]) \cap Q_{j1}^*(S, S')|\\
    &+ |W(\pla{}[Q_{j2}(S, S'), \gamma_2]) \setminus \mc{X}_{j2}| + |W(\pla{}[Q_{j2}(S, S'), \gamma_2]) \cap Q_{j2}^*(S, S')|\\
    &- |W(\pla{}[Q_{j1}(S, S'), \gamma_1]) \cap Q_{j1}^*(S, S') \cap W(\pla{}[Q_{j2}(S, S'), \gamma_2]) \cap Q_{j2}^*(S, S')|\\
    &\geq \IoAgent{}(\plas{}_i)
\end{align}

\noindent
Lastly, since $\IoAgent{}(\plas{}_i)$ is the optimal, the above inequality implies equaliy, that is, 
\begin{equation}
    |S' \cup Z_i(S)| + |W(\pla{}[Q_{j1}(S, S'), \gamma_1]) \cup W(\pla{}[Q_{j2}(S, S'), \gamma_2])| = \IoAgent{}(\plas{}_i)
\end{equation}
This concludes the proof of correctness. As for the running time, one can verify that for each bag $\mc{X}_i$, if $\mc{X}_i$ is of the type \textsf{leaf}, \textsf{forget} or \textsf{introduce}, we need time $O(4^\sigma \cdot n)$ to update all  entries in $H_i$, where $\sigma$ is the treewidth. On the other hand, if $\mc{X}_i$ is of the type \textsf{join}, we need time $O(16^\sigma \cdot n^3)$ to update all entries in $H_i$. Overall, since the number of bags in the tree decomposition is polynomial w.r.t $n$, and $\sigma$ is bounded, the update scheme runs in polynomial time w.r.t. $n$. This concludes the proof.
\end{proof}

\subsection{PTAS on planar graphs}
One can easily verify that \maxinte{}-\IoAgent{} remains hard on planar graphs. Given a planar graph $G$ and for any constant $\epsilon > 0$, we present a {\em polynomial time approximation scheme} that achieves a $(1 - \epsilon)$ approximation for the \maxinte{}-\IoAgent{}. First, based on the algorithm for treewidth bounded graphs, observe that
\begin{mybox2}
\begin{observation}\label{obs:connected-component}
Given a graph $\mc{G}$ such that each connected component is tree-width bounded, the problem \maxinte{}-\IoAgent{} can be solved in polynomial time on $\mc{G}$. 
\end{observation}
\end{mybox2}

\paragraph{PTAS algorithm.} Let $q = 2 \cdot \ceil{1/\epsilon}$. We start with a plane embedding of $\G{}$ which divides the set of vertices into $\ell$ layers. Let $\Gv{}_i$ be the set of vertices in the $i$th layer, $i = 1, ..., \ell$.  For each $r = 1, ..., q$, observe that we may partition the vertex set into $t+1$ subsets, $t = \ceil{(\ell - r) / q}$, such that the $(i)$ the first subset $\mc{W}_{(1, r)}$ consists of the first $r$ layers, $(ii)$ the last subset $\mc{W}_{(t+1, r)}$ consists of the last $\left((l - r) \mod q \right)$ layers, and $(iii)$ each $i$th subset $\mc{W}_{(i, r)}$ in the middle contains $q$ layers in sequential order. Let $\mc{W}_r = \{\mc{W}_{(1, r)}, ..., \mc{W}_{(t+1, r)}\}$ be such a partition. Let $\G{}_{(i,r)}$ be the subgraph induced on $\mc{W}_{(i, r)}$, $i = 1. ,,, t+1$. It is known that each $\G{}_{(i,r)}$ is an outerplanar graph with a bounded treewidth $O(q)$~\cite{bodlaender1998partial}. Let $\G{}_r = \bigcup_i \G{}_{(i,r)}$. Then by Observation~(\ref{obs:connected-component}), we can solve the problem optimally on each $\G{}_r$, $r = 1, ..., q$ in polynomial time. The algorithm then returns the solution with the largest objective over all $r = 1, ..., q$. One can easily verify that the overall scheme runs in polynomial time w.r.t. $n$.

\begin{mybox2}
\begin{theorem}
The algorithms gives a factor $(1 - \epsilon)$ approximation on planar graphs for any fixed $\epsilon > 0$.
\end{theorem}
\end{mybox2}

\begin{proof}
Recall that $k$ is the number of type-\texttt{1} agents (and $n-k$ is the number of type-\texttt{2} agents). Let $q = 2 \cdot \ceil{1/\epsilon}$. We show that the algorithm gives a $1 - 2/q \geq 1 - \epsilon$ approximation. The case for $q < 3$ is trivially true. Let $\mc{P}^*$ be an \assignment{} of agents on $\G{}$ that gives the maximum number of integrated agents. Fix a $r = 1, ..., q$, let $\mc{W}_r = \{\mc{W}_{(1, r)}, ..., \mc{W}_{(t+1, r)}\}$ be a partition of the vertex set as described above. Let $\mc{P}_{r}$ be an \assignment{} on $\G{}_r$ that is obtained from the proposed algorithm. We now look at the \assignment{} $\mc{P}_r$ and $\plas{}$, restricted to vertices in $\mc{W}_r$. Specifically, let $\mc{P}_{(i,r)}$ and $\mc{P}^*_{(i,r)}$ be the \assignment{} of agents restricted to the subset $W_{(i,r)}$ under $\mc{P}_r$ and $\mc{P}^*$, respectively. Further, let $\texttt{IoA}(\mc{P}_{(i,r)})$ be the number of integrated agents in $\G{}_{(i,r)}$ under $\mc{P}_r$, and $\texttt{IoA}(\mc{P}^*_{(i,r)})$ is the number of integrated agents in $\G{}_{(i,r)}$ under $\mc{P}^*$. We first observe that
\begin{equation}\label{eq:ptas-1}
    \texttt{IoA}(\mc{P}_r) = \sum_{i = 1}^{t+1} \texttt{IoA}(\mc{P}_{(i,r)})
\end{equation}
which is true since $\G{}_{(i,r)}$'s are disconnected. Further, by the fact that $\mc{P}_r$ is optimal on $\mc{G}_r$, we have
\begin{equation}\label{eq:ptas-2}
    \sum_{i = 1}^{t+1} \texttt{IoA}(\mc{P}_{(i,r)}) \geq \sum_{i = 1}^{t+1} \texttt{IoA}(\mc{P}^*_{(i,r)})
\end{equation}
Note that $\sum_{i = 1}^{t+1} \texttt{IoA}(\mc{P}^*_{(i,r)})$ could be less than $\texttt{IoA}(\mc{P}^*)$, which is the optimal objective on $\mc{G}$. Let 
$$
 \sum_{i = 1}^{t+1} \texttt{IoA}(\mc{P}^*_{(i,r)}) = \texttt{IoA} (\mc{P}^*) - \Delta_r
$$
where $\Delta_r \geq 0$ is the difference. We note that the integrated vertices that are left uncounted can only exist on the two adjacent layers between each pair of $\G{}_{(i,r)}$ and $\G{}_{(i+1,r)}$, $i = 1,...t$. Let $\mc{V}^*$ be the set of integrated vertices under $\plas{}$. We then have,
\begin{equation}
    \Delta_r \leq \sum_{j = 0}^{t} \left(\mc{V}^* \cap \mc{V}_{j \cdot q + r}\right) + \left(\mc{V}^* \cap \mc{V}_{j \cdot q + r + 1}\right)
\end{equation}
Since the layers are a partition of the vertex set, and each layer gets counted exactly twice in the above sum, We have, 
\begin{equation}
    \sum_{r = 1}^q \Delta_r = 2 \cdot \texttt{IoA} (\mc{P}^*)
\end{equation}

It follows that
\begin{equation}\label{eq:ptas-5}
    \min_{r = 1, ..., q}\{ \Delta_r\} \leq \frac{2}{q} \cdot \texttt{IoA} (\mc{P}^*)
\end{equation}
Let $r^* = \argmin_{r = 1, ..., q}\{ \Delta_r\}$. By equation~(\ref{eq:ptas-1}) and (\ref{eq:ptas-2}), we have 
\begin{equation}\label{eq:ptas-6}
    \IoAgent{}(\mc{P}_{r^*}) \geq (1 - \frac{2}{q}) \cdot  \IoAgent{}(\mc{P}^*)
\end{equation}

Lastly, let $\hat{\mc{P}}$ be the \assignment{} returned by the algorithm, that is, $\hat{\mc{P}} = \argmax_{r} \texttt{IoA}(\mc{P}_r)$. By Equations~(\ref{eq:ptas-1}) to (\ref{eq:ptas-6}), we have
\begin{equation}
    \texttt{IoA}(\hat{\mc{P}}) \geq (1 - \frac{2}{q}) \cdot \texttt{IoA} (\mc{P}^*)
\end{equation}
This concludes the proof. 
\end{proof}


\newpage

\end{document}